\documentclass[fleqn,usenatbib]{mnras}

\usepackage{newtxtext,newtxmath}
\usepackage[T1]{fontenc}

\DeclareRobustCommand{\VAN}[3]{#2}
\let\VANthebibliography\thebibliography
\def\thebibliography{\DeclareRobustCommand{\VAN}[3]{##3}\VANthebibliography}

\usepackage{graphicx}	% Including figure files
\usepackage{amsmath}	% Advanced maths commands
\usepackage{bm}
\usepackage{multirow}
\newcommand{\Lya}{\mathrm{Ly\alpha} }
\newcommand{\nhat}{ \bm{\hat{n}} }
\newcommand{\voigt}{ \mathcal{H} }
\newcommand{\HI}{\mathrm{H\textsc{i}}}

\title[Formulae for $\Lya$ radiative transfer]{Analytical and fitting formulae for solutions to Lyman-alpha radiative transfer equations: the effects of geometry, recoil, and velocity gradients}

\author[Pengfei Li \& Zheng Zheng]{
Pengfei Li\thanks{E-mail: Pengfei.Li@utah.edu}
and Zheng Zheng\thanks{E-mail:
zhengzheng@astro.utah.edu}
\\
Department of Physics and Astronomy, University of Utah, E2108 Stewart Building, 270 S 1400 E, Salt Lake City, UT 84112, USA
}

\date{Accepted XXX. Received YYY; in original form ZZZ}

\pubyear{\the\year{}}

\begin{document}
\label{firstpage}
\pagerange{\pageref{firstpage}--\pageref{lastpage}}
\maketitle

\begin{abstract}
Lyman-alpha ($\Lya$) radiative transfer (RT) is important in many astrophysical environments and governed by multiple physical processes. 
In this paper, we provide analytical formulae/procedures for the solutions to $\Lya$ RT equations under three simple geometrical symmetries and investigate the effects of atomic recoil and gas bulk motion. 
We first study $\Lya$ spectra by solving $\Lya$ RT equations for a static, uniform gas cloud under cylindrical geometry. 
The solution is verified through $\Lya$ Monte Carlo RT simulations, and compared to those under slab and spherical geometries in literature. 
Second, to characterise the recoil effect, we empirically modify recoil-free $\Lya$ spectra. 
The method is motivated by $\Lya$ RT equations with recoil and justified by simulations. 
Finally, we account for constant velocity gradients in $\Lya$ RT equations and obtain series solutions for $\Lya$ spectra. 
The solutions demonstrate good agreement to $\Lya$ spectra from simulations for small velocity gradients (i.e. edge velocity $v_{\rm E}$ of a cloud being comparable to the thermal velocity $b$) but become less accurate for large ones. 
To characterise $\Lya$ spectra under large velocity gradients, we empirically extend the functional form of solutions and constrain them from fitting simulated $\Lya$ spectra. 
The resulting fitting formulae show significant improvement for large velocity gradients ($v_{\rm E}/b \sim 100$) under large optical depths. 
The analytical study of $\Lya$ spectra in this work completes the set of solutions under simple geometries, provides physical insights for $\Lya$ RT under recoil and velocity gradient, and develops analytical tools for theoretical studies that require inputs from $\Lya$ RT. 
\end{abstract}


%%%%%%%%%%%%%%%%%%%%%%%%%%%%%%%%%%%%%%%%%%%%%%%%%%

%%%%%%%%%%%%%%%%% BODY OF PAPER %%%%%%%%%%%%%%%%%%
\section{Introduction}\label{sec:intro}

$\Lya$ emission is one of the strongest emission lines in the Universe \citep{Partridge&Peebles_1967}. 
Detecting $\Lya$ emission has proved fruitful in many aspects, including probing gas inside and around galaxies or quasars \citep[e.g.][]{Bacon_2015}, mapping galaxy distributions in the early Universe \cite[e.g.][]{Gebhardt_2021}, and constraining the epoch of reionization \cite[e.g.][]{Ouchi_2018}.  
The physical interpretation of $\Lya$ emission line profiles or spectra, however, turns out to be theoretically challenging due to the resonant scattering between $\Lya$ photons and neutral hydrogen atoms (i.e. $\Lya$ radiative transfer or $\Lya$ RT). 
$\Lya$ RT process can strongly modify the shape of emission line profiles, shifting the flux peaks, broadening the line widths, and creating asymmetric, multi-peak features. 
In this paper, we develop analytical formulae for $\Lya$ spectra with different physical ingredients that affect $\Lya$ RT, to provide theoretical insights for better characterising $\Lya$ spectral features. 

Many physical factors are involved in $\Lya$ RT. 
The properties of $\HI$ gas clouds, including density, bulk motion, and temperature, jointly determine the $\Lya$ optical depth and subsequently affect the $\Lya$ RT process in the cloud. 
Under large optical depth, the $\Lya$ spectra usually have double-peak features, with the frequency separation between the two peaks primarily dictated by the optical depth \citep[e.g.][]{Unno_1955, Neufeld_1990}. 
Furthermore, the bulk motion of $\HI$ gas strongly modifies the flux ratio of the two peaks \citep[e.g.][]{Zheng_2002a, Verhamme_2006, Song_2020, Garel_2024, Nebrin_2025, Smith_2025}. 
An expanding cloud enhances the flux of the red peak while a contracting cloud enhances that of the blue peak. 
The geometry of a gas cloud also plays a role in shaping $\Lya$ spectra. 
For example, a uniform cloud with slab-like geometry yields broader line profile compared to a spherical cloud \citep[e.g.][]{Dijkstra_2006}.  
A clumpy gas cloud can produce multiple peaks in $\Lya$ spectra, depending on the clump covering factor \citep[e.g.][]{Gronke_2016}. 
Anisotropic gas density distributions can generate anisotropic $\Lya$ emission, resulting in the dependence of $\Lya$ spectral shapes on observing angles \citep[e.g.][]{Zheng_2014, Blaizot_2023}. 
The atomic recoil is another mechanism that modifies $\Lya$ spectra, primarily for gas at low temperature. 
It is relevant for coupling the spin temperature of 21 centimetre and the gas temperature in the early Universe \citep[e.g.][]{Wouthuysen_1952, Field_1959}. 
On average, atomic recoil transfers energy from $\Lya$ photons to $\HI$ atoms, leading to a suppressed blue peak. 
Finally, $\Lya$ photons can be destroyed through collisional de-excitation, orbital-angular-momentum-changing collisions followed by two-photon decay, or absorption by dust or molecular hydrogen \citep[e.g.][]{Pengelly_1964, Harrington_1973, Neufeld_1990, Hansen_2006, Verhamme_2006, Laursen_2013, Tomaselli_2021, Guzman_2017, Nebrin_2025}. 

$\Lya$ RT can be studied through Monte Carlo RT simulations under arbitrary gas configurations. 
However, it is still important to study analytical solutions to $\Lya$ RT equations under simplified configurations for physical insights, which can strongly aid theoretical modelling, statistical inference from observations, and implementation of sub-grid models in galaxy formation simulations. 
Solutions of $\Lya$ spectra under many different physical configurations are presented in literature. 
For example, the solutions to a static, uniform cloud are obtained under slab/cubic geometry \citep[e.g.][]{Unno_1955, Harrington_1973, Neufeld_1990, Tasitsiomi_2006, Lao_2020, Stace_2026} and spherical geometry \citep[e.g.][]{Dijkstra_2006, Lao_2020, McClellan_2022, Nebrin_2025, Smith_2025, Stace_2026}. 
More complicated gas configurations are also accounted for, with density variations \citep[e.g.][]{Lao_2020, Nebrin_2025} and velocity gradients \citep[e.g.][]{Loeb_1999, Nebrin_2025, Smith_2025}. 
Other physical mechanisms are further considered, including dust absorption, $\Lya$ destruction, and recoil \citep[e.g.][]{Harrington_1973, Neufeld_1990, Tomaselli_2021, Nebrin_2025}. 
In particular, the recoil effect is considered for infinite clouds \citep[e.g.][]{Chuzhoy_2006, Furlanetto_2006} and spherical finite clouds \citep{Nebrin_2025}. 

In this paper, we study analytical solutions of $\Lya$ spectra. 
We organize the paper as follows: In Section \ref{sec:cylindrical}, we solve $\Lya$ RT equations for a static, uniform gas cloud with a cylindrical geometry. 
Together with analytical solutions under slab and spherical geometries from literature, it provides the last piece in the set of analytical solutions under simple geometries. 
In Section \ref{sec:recoil}, we incorporate the recoil effect by proposing a simple modification to recoil-free $\Lya$ spectra. 
In Section \ref{sec:vgrad}, we obtain solutions to $\Lya$ RT equations with small velocity gradients, and empirically extend them to describe $\Lya$ spectra under large velocity gradients. 
Comparisons are made among three geometries and between $\Lya$ spectra from analytical formulae and from Monte Carlo $\Lya$ RT simulations in each of the above sections. 
Section \ref{sec:summary} summarizes the main results and discusses further improvements and applications. 
The logarithm $\log$ in this paper is base 10. 
\section{RT solution under cylindrical geometry}\label{sec:cylindrical}
In this section, we derive the $\Lya$ RT equation for a static, uniform gas cloud under cylindrical geometry (with recoil effect neglected) and obtain $\Lya$ spectra by solving it. 
In astrophysical environments such as circum-galactic medium and inter-galactic medium, filamentary structures resemble the cylindrical geometry, and the analytical solution can help us gain insights on $\Lya$ RT effects. 
The solution is verified through $\Lya$ RT simulations and compared to the analytical solutions under slab and spherical geometry in literature. 

For the $\Lya$ RT simulations used in this and following sections, we adopt the Monte Carlo RT method introduced in \cite{Zheng_2002a}, which traces the frequency and spatial evolution of $\Lya$ photons until they escape at the system boundary. The gas temperature is fixed at $T=10\, \rm K$ unless otherwise specified. 
We simulate $10^5$ photons in each run to calculate $\Lya$ spectra. 
No core-skipping approximation is adopted in $\Lya$ RT simulations. 
The gas clouds considered in this paper have large optical depths. 
To enhance simulation efficiency, we improve the sampling algorithm for generating the velocity of the hydrogen atom responsible for the scattering, which is described in Appendix~\ref{sec:randomGenRT}. 

\subsection{RT equation under cylindrical geometry}\label{subsec:RTEquation}
We start from the general RT equation with no time evolution 
\begin{equation}
\begin{split}
    \nhat \cdot \nabla I_\nu(\bm{r}, \nhat) = & -\alpha_\nu (\bm{r})I_\nu(\bm{r}, \nhat) + j_\nu(\bm{r}, \nhat) \\
    & + \int \frac{{\rm d} \Omega^\prime}{4\pi} \int {\rm d} \nu^\prime \alpha_{\nu^\prime} I_{\nu^\prime}(\bm{r}, \nhat^\prime) R_{\rm II}(\nu, \nu^\prime, \nhat, \nhat^\prime), 
    \label{eq:RTEGeneral}
\end{split}
\end{equation}
where $I_\nu(\bm{r}, \nhat)$ is the intensity of photons at frequency $\nu$ propagating towards direction $\nhat$ at position $\bm{r}$, $\alpha_\nu$ is the absorption coefficient, $j_\nu$ is the emissivity per solid angle, and $R_{\rm II}(\nu, \nu^\prime, \nhat, \nhat^\prime)$ is the $\Lya$ redistribution function which describes the frequency and direction redistribution of photons due to scattering \citep[e.g.][]{Unno_1952, Hummer_1962}. 

In a static uniform cloud, the absorption coefficient can be written as $\alpha_\nu=n_0\sigma_0\voigt(x)$, where $n_0$ is the $\HI$ gas density, $\sigma_0=5.88 \times 10^{-13} ( T/100\, {\rm K})^{-1/2}\,{\rm cm^2}$ is the cross-section at the $\Lya$ line centre, and $\voigt(x)$ is the Voigt profile \citep[e.g.][]{Dijkstra_2014}  
\begin{equation}
\begin{split}
    \voigt (x) = & \frac{a_{\rm v}}{\pi}\int_{-\infty}^{+\infty} {\rm d}u \frac{e^{-u^2}}{(x-u)^2 + a_{\rm v}^2} \\
    \simeq &
    \begin{cases}
    e^{-x^2} & |x| \ll 1\\
    \frac{a_{\rm v}}{\sqrt{\pi}x^2} & |x| \gg 1. 
    \end{cases}
\end{split}
\label{eq:voigt}
\end{equation}
The parameter $a_{\rm v}=4.7 \times 10^{-3} ( T/100\, {\rm K})^{-1/2}$ is the Voigt parameter. 
The frequency parameter $x$ is defined as $(\nu - \nu_0)/\Delta \nu_{\rm D}$, where $\nu_0$ is the $\Lya$ frequency at the line centre and $\Delta \nu_{\rm D}$ is the Doppler width $ \nu_0 b/c$ at temperature $T$, with thermal velocity $b=\sqrt{2k_{\rm B}T/m_{\rm H}}$, the Boltzmann constant $k_{\rm B}$, the atomic hydrogen mass $m_{\rm H}$, and the speed of light $c$. 
In our definition, the Voigt profile has normalization $\int_{-\infty}^{+\infty} {\rm d}x \voigt(x) = \sqrt{\pi}$.  

First by assuming large optical depth $a_{\rm v} \tau_0 \gtrsim 10^3$ \citep{HummerKunasz_1980, Neufeld_1990, Nebrin_2025, Lorinc_2025}, the scattering mostly happens in the line wings, where $\voigt(x) \simeq a_{\rm v}/ (\sqrt{\pi}x^2 )$.\footnote{The value $\sim 10^3$ is obtained in \cite{HummerKunasz_1980} by examining the scaling relation between the average number of scatterings and the optical depth. 
They found that the scaling relation predicted by the analytical solution from \cite{Harrington_1973} converges to that from their numerical solution when $a_{\rm v} \tau_0 \gtrsim 5 \times 10^3$. }
Then by applying the Fokker-Planck approach \citep[e.g.][]{Unno_1955, Harrington_1973, Rybicki_1994, Rybicki_2006}, the integral term on the right-hand side of equation~(\ref{eq:RTEGeneral}) can be approximated by 
\begin{equation}
     n_0 \sigma_0 \left[\voigt J + \frac{1}{2} \frac{\partial }{\partial x}\left( \voigt \frac{\partial J}{\partial x} \right)\right], 
    \label{eq:RApprox}
\end{equation}
where $J \equiv \int {\rm d} \Omega\, I_\nu/(4\pi)$ is the zeroth moment of the intensity (i.e. mean intensity). 
With these steps, equation~(\ref{eq:RTEGeneral}) becomes 
\begin{equation}
\begin{split}
      \frac{\nhat \cdot \nabla I_\nu}{n_0 \sigma_0} = -\voigt I_\nu + \frac{j_\nu}{n_0 \sigma_0}  
    + \voigt J + \frac{1}{2} \frac{\partial }{\partial x}\left( \voigt \frac{\partial J}{\partial x} \right).
    \label{eq:RTEG_cylindrical}
\end{split}
\end{equation}
We evaluate the zeroth and first angular moments of equation~(\ref{eq:RTEG_cylindrical}) assuming isotropic emissivity $j_\nu$, which yields \citep[e.g.][]{Nebrin_2025}
\begin{equation}
      \frac{\nabla \cdot \bm{H}}{n_0 \sigma_0} = 
      \frac{j_\nu}{n_0 \sigma_0} +  \frac{1}{2} \frac{\partial }{\partial x}\left( \voigt \frac{\partial J}{\partial x} \right), 
    \label{eq:RTEG_moment1}
\end{equation}
and 
\begin{equation}
     \frac{\nabla \cdot \mathbf{K}}{n_0 \sigma_0} = -\voigt \bm{H}.
    \label{eq:RTEG_moment2}
\end{equation}
The first and second angular moments of intensity $I_\nu$ are denoted as $\bm{H} \equiv \int [{\rm d} \Omega/(4\pi)]\, \nhat I_\nu$ and $\mathbf{K} \equiv \int [{\rm d} \Omega/(4\pi)] \, \nhat \nhat I_\nu$, respectively, with $\nhat \nhat$ being the dyadic product of two $\nhat$. 
Taking the divergence of equation~(\ref{eq:RTEG_moment2}) and substituting it to equation~(\ref{eq:RTEG_moment1}), we arrive at 
\begin{equation}
    \frac{\nabla \cdot (\nabla \cdot \mathbf{K})}{(n_0 \sigma_0)^2} = 
      -\frac{\voigt j_\nu}{n_0 \sigma_0} - \frac{1}{2} \voigt \frac{\partial }{\partial x}\left( \voigt \frac{\partial J}{\partial x} \right), 
    \label{eq:RTEG_KJ}
\end{equation}
where $\nabla \cdot (\nabla \cdot \mathbf{K}) = \sum_{ij}\partial_i \partial_j K_{ij}$. 

Now, we carry out the Eddington approximation to equation~(\ref{eq:RTEG_KJ}) to obtain a closed equation for $J$, assuming that the intensity $I_\nu$ only has a linear dependence on the cosine angle between propagation direction $\nhat$ and unit position vector $\bm{\hat{r}}$ (i.e. $\nhat \cdot \bm{\hat{r}}$). 
This leads to the closure relation $\mathbf{K} = \bm{1}J/3$ ($\bm{1}$ is a $3\times 3$ identity matrix), which reduces equation~(\ref{eq:RTEG_KJ}) to 
\begin{equation}
    \frac{\nabla^2 J}{(n_0 \sigma_0)^2} + \frac{3}{2} \voigt \frac{\partial }{\partial x}\left( \voigt \frac{\partial J}{\partial x} \right) = -\frac{3\voigt j_\nu}{n_0 \sigma_0} . 
    \label{eq:RTEG_Laplace_J}
\end{equation}
Under cylindrical symmetry, the mean intensity $J$ only depends on the cylindrical radial coordinate. 
By expressing the Laplace operator $\nabla^2$ under cylindrical coordinate system, equation~(\ref{eq:RTEG_Laplace_J}) transforms into 
\begin{equation}
    \frac{\partial^2 J}{\partial \tau^2} + \frac{1}{\tau}\frac{\partial J}{\partial \tau} + \frac{3}{2} \voigt \frac{\partial}{\partial x}\left( \voigt \frac{\partial J}{\partial x} \right)= -\frac{3\voigt j_\nu}{n_0\sigma_0}, 
    \label{eq:RTE_cylindricalx}
\end{equation}
where the optical depth parameter $\tau$ is defined following ${\rm d}\tau = n_0 \sigma_0 {\rm d} r$ with $r$ the radial distance in the cylindrical coordinate system\footnote{The meaning of $r$ depends on geometry in this paper.}.
We introduce another frequency parameter $y(x)$ defined as \citep[e.g.][]{Nebrin_2025} 
\begin{equation}
    y(x) = \sqrt{\frac{2}{3}}\int_0^x \frac{{\rm d} u}{\voigt(u)} 
    \simeq 
    \begin{cases}
    \sqrt{\frac{2}{3}} x  & |x| \ll 1\\
    \sqrt{\frac{2\pi}{27}} \frac{x^3}{a_{\rm v}} & |x| \gg 1,
    \end{cases}
    \label{eq:yx}
\end{equation}
with which equation~(\ref{eq:RTE_cylindricalx}) can be rewritten as 
\begin{equation}
    \frac{\partial^2 J}{\partial \tau^2} + \frac{1}{\tau}\frac{\partial J}{\partial \tau} + \frac{\partial^2 J}{\partial y^2}= -\frac{3\voigt j_\nu}{n_0\sigma_0}. 
    \label{eq:RTE_cylindrical}
\end{equation}

\subsection{Solution under cylindrical geometry} \label{subsec:ses_cyl}
Equation~(\ref{eq:RTE_cylindrical}) can be solved using an eigenfunction expansion following previous work \citep[e.g.][]{Harrington_1973, Neufeld_1990, Dijkstra_2006} 
\begin{equation}
    J(y,\tau) = \sum_{n=0}^{\infty} E_n(\tau)h_n(y), 
    \label{eq:ses_expansion}
\end{equation}
where $E_n(\tau)$ satisfies the eigenfunction equation with $\lambda_n$ being the eigenvalue 
\begin{equation}
    \frac{{\rm d^2} E_n}{{\rm d}\tau^2} + \frac{1}{\tau} \frac{{\rm d} E_n}{{\rm d}\tau} + \lambda_n^2 E_n = 0, \,\,\,\,\,\,\, n=0, 1, 2...
    \label{eq:ses_En}
\end{equation} 
This is the Bessel equation in the parametrized form, which has the general solution 
\begin{equation}
    E_n(\tau) = AJ_0(\lambda_n\tau) + BY_0(\lambda_n \tau), 
    \label{eq:En_general}
\end{equation}
with $J_0$ and $Y_0$ being the zeroth-order Bessel functions of the first and second kinds, respectively, and with $A$ and $B$ being the coefficients to be determined. 
The solution should converge at $\tau \to 0$, which sets $B=0$. 

The parameter $\lambda_n$ is determined at the boundary $r=r_0$, or equivalently $\tau=\tau_0 \equiv n_0 \sigma_0 r_0$. 
The intensity $I_\nu$ at the boundary $\tau_0$ is assumed to have no angular dependence and only travel outwards, which leads to $\bm{H}=\hat{\bm{r}} J/2$. 
With the Eddington approximation $\mathbf{K} = \bm{1}J/3$ applied to equation~(\ref{eq:RTEG_moment2}), the boundary condition is derived to be \citep[e.g.][]{Harrington_1973, Neufeld_1990}
\begin{equation}
    \frac{\partial J(\tau_0)}{\partial \tau} =-\frac{3}{2} \voigt J(\tau_0). 
    \label{eq:bc}
\end{equation}
Applying equation~(\ref{eq:bc}) to equation~(\ref{eq:En_general}), we obtain 
\begin{equation}
    \lambda_n  J_1(\lambda_n \tau_0) = \frac{3}{2} \voigt J_0(\lambda_n\tau_0),
    \label{eq:bc_cyl}
\end{equation}
where $J_1$ is the first-order Bessel function of the first kind. 
The Bessel functions $J_0(z)$ and $J_1(z)$ have asymptotic forms for large $z$
\begin{equation}
    J_0(z) \simeq \sqrt{\frac{2}{\pi z}} \cos(z - \frac{\pi}{4}),\,\,\,\,\,\, z \gtrsim 1
    \label{eq:J0_approx}
\end{equation}
and 
\begin{equation}
    J_1(z) \simeq \sqrt{\frac{2}{\pi z}} \sin(z - \frac{\pi}{4}),\,\,\,\,\,\, z \gtrsim 2.
    \label{eq:J1_approx}
\end{equation}
Applying equation~(\ref{eq:J0_approx}) and (\ref{eq:J1_approx}) to equation~(\ref{eq:bc_cyl}), we obtain 
\begin{equation}
    \lambda_n \tan(\lambda_n \tau_0 - \frac{\pi}{4}) = \frac{3}{2}\voigt \Rightarrow \lambda_n \tau_0 = n\pi + \frac{\pi}{4} + \tan^{-1} \left( \frac{3\voigt}{2\lambda_n} \right).
    \label{eq:bc_approx}
\end{equation}
For large optical depths, the term $\tan^{-1} \left( 3\voigt/2\lambda_n \right)$ approaches $\pi/2$, which sets $\lambda_n$ to be\footnote{We notice that the term $\tan^{-1} \left( 3\voigt/2\lambda_n \right)$ can not be approximated as $\pi/2$ when $n \to \infty$. 
However, the contribution from higher-order terms with large $n$ is exponentially suppressed as to be shown in equation~(\ref{eq:hn_general}). }
\begin{equation}
    \lambda_n = \frac{1}{\tau_0}(n+ \frac{3}{4})\pi, \,\,\,\,\,\, n=0,1,2,....
    \label{eq:bc_lambdan}
\end{equation}

With equation~(\ref{eq:bc_lambdan}) and equation~(\ref{eq:J0_approx}), it is straightforward to see that function $J_0(\lambda_n \tau)$ is zero at boundary $\tau = \tau_0$, which means that $J_0(\lambda_n\tau)$ forms an orthogonal basis, such that 
\begin{equation}
    \int_0^{\tau_0} \tau J_0(\lambda_m \tau)J_0(\lambda_n \tau){\rm d} \tau=\frac{\tau_0^2}{2}J_1^2(\lambda_n\tau_0)\delta_{mn} \simeq \frac{\tau_0}{\pi \lambda_n} \delta_{mn}, 
    \label{eq:J0_cob}
\end{equation}
where $\delta_{mn}$ is the Kronecker delta function, and the second step evaluates $J_1(\lambda_n \tau_0)$ using equation~(\ref{eq:J1_approx}). 
We fix the coefficient $A$ in equation~(\ref{eq:En_general}) requiring $\int_0^{\tau_0} \tau E_n^2(\tau) {\rm d} \tau=1$, which sets $A= \sqrt{2}/(\tau_0 |J_1(\lambda_n \tau_0)|) \simeq \sqrt{\pi \lambda_n/\tau_0}$. 
The eigenfunction $E_n$ now reads
\begin{equation}
    E_n(\tau)=\frac{\sqrt{2}J_0(\lambda_n\tau)}{\tau_0|J_1(\lambda_n\tau_0)|}.
    \label{eqn:En}
\end{equation}
By substituting equation~(\ref{eq:ses_expansion}) into equation~(\ref{eq:RTE_cylindrical}) and applying the orthogonal relation of $E_n$, we obtain the equation for $h_n(y)$, 
\begin{equation}
    \frac{{\rm d}^2 h_n}{{\rm d} y^2} - \lambda_n^2 h_n = -\frac{3\voigt Q_n}{n_0 \sigma_0},
    \label{eq:ses_hn}
\end{equation}
where $Q_n = \int_0^{\tau_0} \tau j_\nu E_n d\tau$ is the source term. 

We consider a cylindrical source emitting isotropically at radius $r_{\rm s}$ at a single frequency $x_{\rm i}$, 
\begin{equation}
    j_\nu = \frac{1}{4\pi}\frac{\delta(r-r_{\rm s})}{2\pi r_{\rm s}}  \delta(x - x_{\rm i})
          = \frac{n_0^2 \sigma_0^2}{4\sqrt{6} \pi^2 \tau_{\rm s} \voigt(y_{\rm i})} \delta(\tau-\tau_{\rm s})\delta(y - y_{\rm i}), 
    \label{eq:jnu}
\end{equation}
where $\delta$ is the Dirac delta function, $\tau_{\rm s} \equiv n_0\sigma_0r_{\rm s}$, and $y_{\rm i} = \sqrt{2/3}\int_0^{x_{\rm i}}{\rm d}u /\voigt(u)$ following equation~(\ref{eq:yx}). 
The emissivity $j_\nu$ satisfies normalization $\int {\rm d}\Omega\, {\rm d}r\, {\rm d}x\, (2\pi r j_\nu) =1$ under our definition\footnote{In our setup, the source only emits at a single frequency $x_{\rm i}$ in the lab frame, and does not follow a Voigt profile as assumed in previous literatures \citep{Harrington_1973, Nebrin_2025}. 
We argue that this is necessary to analytically carry on the calculation when $x_{\rm i} \neq 0$. 
By assuming $j_\nu \propto \voigt$, previous work obtains a $\voigt^2(x)$ term from the right-hand side of equation~(\ref{eq:ses_hn}), and proceeds the calculation through approximation $\voigt^2(x) \simeq \sqrt{2\pi/3}\delta(y)$ \citep[e.g.][]{Harrington_1973}. 
This is only equivalent to our setup when $x_{\rm i}=0$. 
While for $x_{\rm i} \neq 0$, the $\voigt^2$ term becomes $\voigt(x)\voigt(x-x_{\rm i})$, and can not be approximated as $\sqrt{2\pi/3}\delta(y - y_{\rm i})$. 
With our assumption of $j_\nu$, we avoid this problem and keep the calculation analytically tractable. 
The situation where $x_{\rm i} \neq 0$ has already been studied in \cite{Neufeld_1990}, who obtained consistent results with ours but made an unnatural assumption of the emissivity in their appendix A. 
}.
As a consequence, both the emissivity $j_\nu$ and the mean intensity $J$ have a different dimension than the normal definition. 
With $j_\nu$ in equation~(\ref{eq:jnu}), the source term $Q_n$ becomes 
\begin{equation}
    Q_n = \frac{n_0^2 \sigma_0^2}{4\sqrt{3} \pi^2 \tau_0 \voigt(y_{\rm i})} \frac{J_0(\lambda_n \tau_{\rm s})}{|J_1(\lambda_n \tau_0)|} \delta(y-y_{\rm i}), 
    \label{eq:Qn}
\end{equation}
which is zero everywhere except at $y = y_{\rm i}$. 
Plugging $Q_n$ into equation~(\ref{eq:ses_hn}), the solution can be written as 
\begin{equation}
    h_n(y) = C\exp(-\lambda_n |y - y_{\rm i}|). 
    \label{eq:hn_general}
\end{equation}
By integrating equation~(\ref{eq:ses_hn}) near $y_{\rm i}$ from $y_{\rm i}^-$ to $y_{\rm i}^+$ and taking the limit of $y_{\rm i}^+ = y_{\rm i}^- = y_{\rm i}$, the coefficient $C$ can be determined by the jump-condition of the first derivative of $h_n$ at $y = y_{\rm i}$, 
\begin{equation}
    \left. \frac{{\rm d} h_n}{{\rm d} y}\right|_{y_{\rm i}^+} - \left.\frac{{\rm d} h_n}{{\rm d} y}\right|_{y_{\rm i}^-} = -2\lambda_n C=-\frac{\sqrt{3}n_0 \sigma_0}{4\pi^2 \tau_0} \frac{J_0(\lambda_n \tau_{\rm s})}{|J_1(\lambda_n \tau_0)|} .
    \label{eq:jump}
\end{equation}

Combining equation~(\ref{eq:bc_lambdan}), (\ref{eqn:En}), and (\ref{eq:hn_general}), we obtain the series solution to $\Lya$ RT equation~(\ref{eq:RTE_cylindrical}) for a cylindrical source located at radius $r_{\rm s}$ emitting at a frequency $x_{\rm i}$ under cylindrical geometry 
\begin{equation}
    J(y, \tau) = \frac{\sqrt{6}n_0 \sigma_0}{8\pi^2 \tau_0^2} \sum_{n=0}^{\infty}\frac{J_0(\lambda_n \tau_{\rm s})}{\lambda_n J_1^2(\lambda_n \tau_0)}J_0(\lambda_n\tau) \exp(-\lambda_n |y - y_{\rm i}|).
    \label{eq:ses_full_tau}
\end{equation}
To calculate the emergent spectrum at the boundary $\tau=\tau_0$, we apply equation~(\ref{eq:bc_cyl}) to substitute $J_0(\lambda_n\tau_0)$ and obtain 
\begin{equation}
    J_{\rm ses}(y) = \frac{\sqrt{6}n_0 \sigma_0}{12\pi^2 \tau_0^2 \voigt} \sum_{n=0}^{\infty}\frac{J_0(\lambda_n \tau_{\rm s})}{J_1(\lambda_n \tau_0)}\exp(-\lambda_n |y - y_{\rm i}|). 
    \label{eq:ses_full}
\end{equation}
Applying the asymptotic form of Bessel functions, equation~(\ref{eq:J0_approx}) and (\ref{eq:J1_approx}), to the emergent spectrum, equation~(\ref{eq:ses_full}), and making use of the Euler equation $e^{iz} = \cos z + i \sin z$, we find a closed functional form of the series in equation~(\ref{eq:ses_full}) as
\begin{equation}
\begin{split}
    J_{\rm cls}(y) = & \frac{\sqrt{6}n_0 \sigma_0}{24 \pi^2 \tau_0^{3/2} \tau_{\rm s}^{1/2} \voigt} 
           \times   \frac{1}{\cos \left( \frac{\pi\tau_{\rm s}}{\tau_0} \right) + \cosh \left( \frac{\pi |y - y_{\rm i}|}{\tau_0} \right)} \\
    \times & \left[ \exp \left( \frac{\pi |y - y_{\rm i}|}{4\tau_0} \right) \cos \left( \frac{3\pi\tau_{\rm s}}{4\tau_0} - \frac{\pi}{4} \right) \right. \\ 
     & \left. + \exp \left( -\frac{3\pi |y - y_{\rm i}|}{4\tau_0} \right) \cos \left( \frac{\pi\tau_{\rm s}}{4\tau_0} + \frac{\pi}{4}\right) \right].
    \label{eq:ana_aspt}
\end{split}
\end{equation}
However, equation~(\ref{eq:ana_aspt}) does not faithfully reproduce equation~(\ref{eq:ses_full}) due to the difference between Bessel functions and their asymptotic forms. 
In particular, we notice that equation~(\ref{eq:ana_aspt}) does not converge when $\tau_{\rm s} \to 0$ due to the divergence of equation~(\ref{eq:J0_approx}) as $z \to 0$. 

\begin{figure}
    \includegraphics[alt={A graph showing the ratio of the series solution to that of the closed-form solution for cylindrical geometry with various source positions. A fitting formula is developed to characterise the ratio, which demonstrates good performance with only per cent level differences.}, width=\columnwidth]{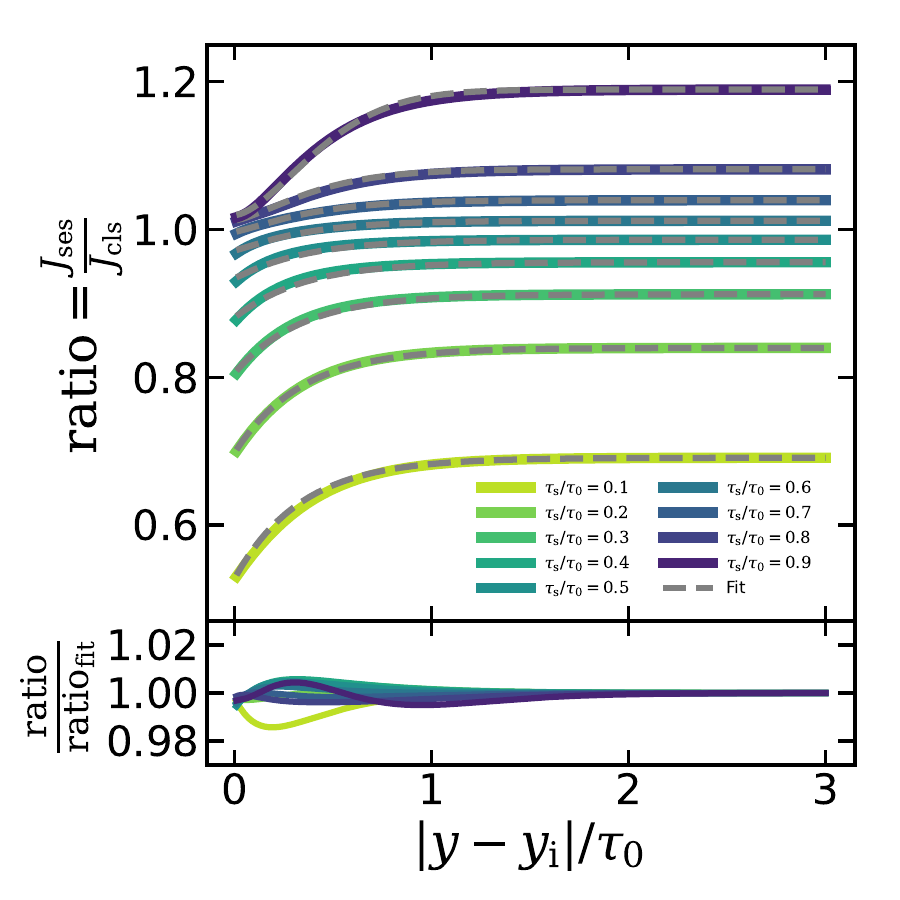}
    \caption{Ratio $J_{\rm ses}/J_{\rm cls}$ of the series solution in equation~(\ref{eq:ses_full}) to the closed-form solution in equation~(\ref{eq:ana_aspt}) as a function of the frequency parameter for different source positions $\tau_{\rm s}/\tau_0$. 
    The dashed lines are the fitting results from equation~(\ref{eq:ratiofit_full}). 
    The bottom panel compares the ratio from $J_{\rm ses}/J_{\rm cls}$ to that from the fitting formula, which demonstrates good agreement with only per cent level differences.  
    }
    \label{fig:cyl_ratiofit}
\end{figure}

As a comparison between the series solution in equation~(\ref{eq:ses_full}) and the closed-form solution in equation~(\ref{eq:ana_aspt}), we plot the ratio of them $R_{\rm sc} = J_{\rm ses}/J_{\rm cls}$ as a function of $|y-y_{\rm i}|/\tau_0$ for different $\tau_{\rm s}/\tau_0$ in Fig.~\ref{fig:cyl_ratiofit}. 
The closed-form solution best mimics the series solution at $\tau_{\rm s}/\tau_0 \sim 0.6$ and becomes less accurate towards either small or large $\tau_{\rm s}/\tau_0$. 

To improve the accuracy of the closed-form solution, we attempt to find a fitting formula for the ratio $R_{\rm sc}$. 
First, when $|y-y_{\rm i}|/\tau_0$ is large, the summation in equation~(\ref{eq:ses_full}) is dominated by the first term ($n=0$) in the series, such that $J(y) \propto J_0(\lambda_0 \tau_{\rm s})/J_1(\lambda_0 \tau_0) \simeq \sqrt{\tau_0/\tau_{\rm s}} \cos(\lambda_0 \tau_{\rm s}-\pi/4)/\sin(\lambda_0 \tau_0 -\pi/4)$. 
In this case, the ratio $R_{\rm sc}$ is simply a function of $\tau_{\rm s}/\tau_0$, given by 
\begin{equation}
    R_{\rm sc, L} = \left[ \frac{J_0(\lambda_0 \tau_{\rm s})}{J_1(\lambda_0 \tau_0)} \right] 
    \left/ \left[ \sqrt{\frac{\tau_0}{\tau_{\rm s}}} \frac{\cos(\lambda_0 \tau_{\rm s}-\pi/4)}{\sin(\lambda_0 \tau_0 -\pi/4)} \right] \right. ,\, \frac{|y-y_{\rm i}|}{\tau_0} \gg 1 .
    \label{eq:ratiofit_largey}
\end{equation} 
Next, when $|y-y_{\rm i}|/\tau_0 \to 0$, we find that $R_{\rm sc}$ can be approximated by power-law functions of $\tau_{\rm s}/\tau_0$, 
\begin{equation}
    R_{\rm sc, S} = 1.86 (\tau_{\rm s}/\tau_0)^{0.5} - 0.84(\tau_{\rm s}/\tau_0)^{1.14}, \,\,\,\,\,\, \frac{|y-y_{\rm i}|}{\tau_0} \ll 1 .
    \label{eq:ratiofit_smally}
\end{equation}
Finally, we assemble equation~(\ref{eq:ratiofit_largey}) and (\ref{eq:ratiofit_smally}) utilizing the following function 
\begin{equation}
    R_{\rm sc,fit} = P(y) R_{\rm sc, S} + [1-P(y)] R_{\rm sc, L}, 
    \label{eq:ratiofit_full}
\end{equation}
where function $P(y)$ is found to be 
\begin{equation}
    P(y) = \exp \left[ -\pi  ( |y-y_{\rm i}|/\tau_0 )^{1+(\tau_{\rm s}/\tau_0)^{3.61}}  \right]. 
    \label{eq:ratiofit_middley}
\end{equation}
The functional forms and parameters in equation~(\ref{eq:ratiofit_smally}), (\ref{eq:ratiofit_full}), and (\ref{eq:ratiofit_middley}) are empirically determined to reproduce $R_{\rm sc}$, and can be modified if better performance is desired. 

In the top panel of Fig.~\ref{fig:cyl_ratiofit}, we plot the fitting results for the ratio $R_{\rm sc,fit}$ in grey dashed lines. It agrees well with the true ratio $R_{\rm sc}$, with only per cent level difference as shown by the bottom panel. 
By multiplying this fitting formula of $R_{\rm sc,fit}$ to equation~(\ref{eq:ana_aspt}), an accurate closed-form solution is obtained as $J_{\rm cyl} = J_{\rm cls}R_{\rm sc,fit}$, whose full functional form is  
\begin{equation}
\begin{split}
    J_{\rm cyl}(y) = & \frac{1}{2\pi r_0} \frac{\sqrt{6}x^2}{12 \sqrt{\pi} a_{\rm v}\tau_0} 
           \frac{1}{\cos \left( \frac{\pi\tau_{\rm s}}{\tau_0} \right) + \cosh \left( \sqrt{ \frac{2\pi^3}{27}} \frac{ \left|x^3 - x_{\rm i}^3\right|}{a_{\rm v}\tau_0} \right)} \\
    \times & \left[ \exp \left( \frac{1}{4} \sqrt{ \frac{2\pi^3}{27}} \frac{ \left|x^3 - x_{\rm i}^3\right|}{a_{\rm v}\tau_0} \right) \cos \left( \frac{3\pi\tau_{\rm s}}{4\tau_0} - \frac{\pi}{4} \right) \right. \\ 
     & \left. + \exp \left( -\frac{3}{4}\sqrt{ \frac{2\pi^3}{27}} \frac{ \left|x^3 - x_{\rm i}^3\right|}{a_{\rm v}\tau_0} \right) \cos \left( \frac{\pi\tau_{\rm s}}{4\tau_0} + \frac{\pi}{4}\right) \right] \\
     \times & \left\{ \exp \left[ -\pi  \left( \sqrt{ \frac{2\pi}{27}} \frac{ \left|x^3 - x_{\rm i}^3\right|}{a_{\rm v}\tau_0} \right)^{1+(\tau_{\rm s}/\tau_0)^{3.61}}  \right] \right. \\
     & \times \left[ 1.86 - 0.84(\tau_{\rm s}/\tau_0)^{0.64} \right] \\
     & + \left\{ 1- \exp \left[ -\pi  \left( \sqrt{ \frac{2\pi}{27}} \frac{ \left|x^3 - x_{\rm i}^3\right|}{a_{\rm v}\tau_0} \right)^{1+(\tau_{\rm s}/\tau_0)^{3.61}}  \right] \right\} \\
     & \left. \times \left[ \frac{J_0 \left( \frac{3\pi}{4} \frac{\tau_{\rm s}}{\tau_0} \right)}{J_1 \left(\frac{3\pi}{4} \right) \cos \left( \frac{3\pi}{4} \frac{\tau_{\rm s}}{\tau_0}-\frac{\pi}{4} \right) } \right] \right\}. 
    \label{eq:cyl_full}
\end{split}
\end{equation}
The Voigt function $\voigt$ is substituted by $a_{\rm v}/(\sqrt{\pi}x^2)$ under large optical depths following equation~(\ref{eq:voigt}). 
The parameters $y$ and $y_{\rm i}$ are expressed in terms of $x$ and $x_{\rm i}$ following equation~(\ref{eq:yx}) under $|x| \gg 1$. 
Compared to equation~(\ref{eq:ana_aspt}), the new functional form not only provides a more accurate evaluation for different $\tau_{\rm s}/\tau_0$, but also fixes the diverging problem as $\tau_{\rm s} \to 0$. 

The middle row of Fig.~\ref{fig:spec_3param} shows the comparison between $\Lya$ spectra from the analytical solution $J_{\rm cyl}$ (dashed lines) and from radiative transfer (RT) simulations (solid lines), for different total optical depths $\tau_0$, source optical depth $\tau_{\rm s}$, and initial frequencies $x_{\rm i}$ in the three columns, respectively. 
The analytical solutions agree well with RT simulations for most frequencies except at peak positions. 
The agreement is reasonable even for cases where $a_{\rm v}\tau_0$ is around $10^2$ that does not satisfy $a_{\rm v}\tau_0 \gtrsim 10^3$ as assumed, which indicates a broader applicable range of the analytical solutions. 

The analytic form of the solution in equation~(\ref{eq:cyl_full}) is not unique. In Appendix~\ref{sec:spec_additional_derivation}, we provide an alternative form of the solution based on a different derivation. 

\subsection{Comparisons to solutions under slab and spherical geometries}\label{subsec:ses_slasph}

\begin{figure*}
    \includegraphics[alt={A comparison between Lyman-alpha spectra from closed-form solutions and from Lyman-alpha radiative transfer simulations for various optical depths, source positions, initial frequencies, and geometries. The closed-form solutions agree well with Lyman-alpha radiative transfer simulations.}, width=\textwidth]{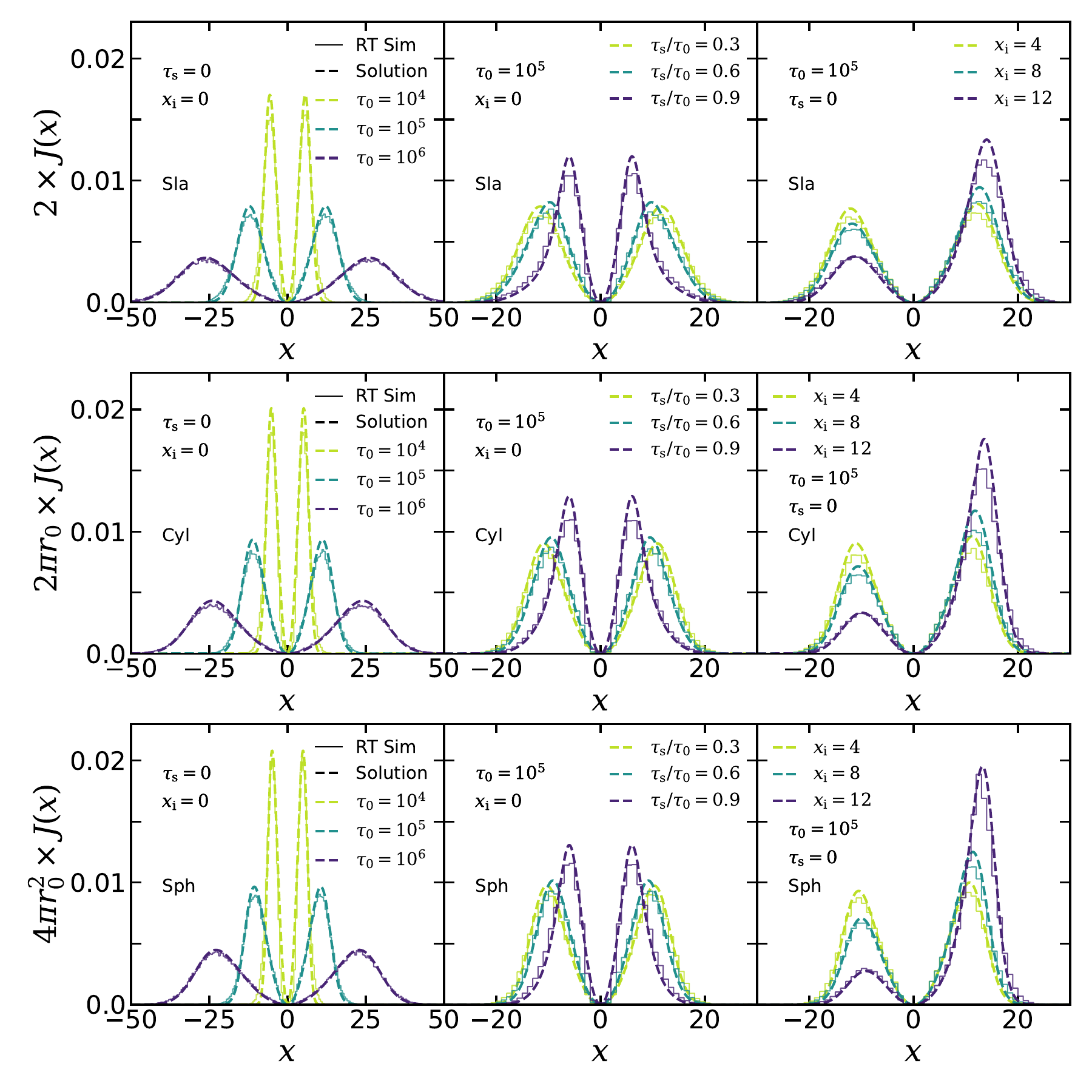}
    \caption{Comparisons between $\Lya$ spectra from closed-form solutions (dashed lines; equation~(\ref{eq:pla_full}), (\ref{eq:cyl_full}), and (\ref{eq:sph_full})) and $\Lya$ RT simulations (solid lines) for various optical depths $\tau_0$, source positions $\tau_{\rm s}$, initial frequencies $x_{\rm i}$, and geometries. 
    The three rows correspond to slab, cylindrical, and spherical geometries, respectively. 
    The three columns correspond to varying one of the three parameters, $\tau_0$, $\tau_{\rm s}/\tau_0$, and $x_{\rm i}$, respectively, with the rest two parameters fixed as labelled in each panel. 
    }
    \label{fig:spec_3param}
\end{figure*}

The $\Lya$ spectra from a static uniform cloud under slab and spherical geometries have been extensively studied in previous literatures. 
In this section, we simply list some key steps for obtaining such solutions, and compare them to the solution we derive for the cylindrical geometry. 

For the slab geometry, the cloud extends from $-r_0$ to $r_0$ in one dimension and infinitely in the rest two dimensions. 
The optical depth $\tau$ is evaluated only along the finite dimension with the cloud centre being $r=0$. 
With such a configuration, the RT equation can be written as 
\begin{equation}
    \frac{\partial^2 J}{\partial \tau^2} + \frac{\partial^2 J}{\partial y^2}= -\frac{3\voigt j_\nu}{n_0\sigma_0}, 
    \label{eq:RTE_plane_parallel}
\end{equation}
where the isotropic source function $j_\nu$ is defined as 
\begin{equation}
    j_\nu = \frac{1}{4\pi}\frac{\delta(|r|-r_{\rm s})}{2}  \delta(x - x_{\rm i}). 
    \label{eq:jnu_pla}
\end{equation}
We require the source position to be symmetric about the cloud centre at $r=\pm r_{\rm s}$, for better comparison with the other two symmetries. 
The source function satisfies normalisation $\int {\rm d}\Omega\, {\rm d}r\, {\rm d}x\, j_\nu =1$ under our definition. 
The boundary condition still satisfies equation~(\ref{eq:bc}). 
With the setup above, the closed-form solution for slab geometry is 
\begin{equation}
\begin{split}
    J_{\rm sla} = \frac{1}{2}\frac{\sqrt{6}x^2}{12 \sqrt{\pi} a_{\rm v}\tau_0}  \frac{ 2\cosh \left( \frac{1}{2} \sqrt{ \frac{2\pi^3}{27}} \frac{ \left|x^3 - x_{\rm i}^3 \right|}{a_{\rm v}\tau_0} \right) \cos\left(\frac{\pi\tau_{\rm s}}{2\tau_0}\right) }{\cos\left( \frac{\pi\tau_{\rm s}}{\tau_0} \right) + \cosh \left( \sqrt{ \frac{2\pi^3}{27}} \frac{ \left|x^3 - x_{\rm i}^3 \right|}{a_{\rm v}\tau_0} \right)}. 
    \label{eq:pla_full}
\end{split}
\end{equation}
This solution is a special case of that in \cite{Neufeld_1990} by requiring symmetric source distribution with respect to the slab centre. 

For the spherical geometry, the gas cloud has finite size $r_0$. 
The optical depth $\tau$ is evaluated along the radial direction from the cloud centre $r=0$. 
The RT equation yields 
\begin{equation}
    \frac{\partial^2 J}{\partial \tau^2} + \frac{2}{\tau}\frac{\partial J}{\partial \tau} + \frac{\partial^2 J}{\partial y^2}= -\frac{3\voigt j_\nu}{n_0\sigma_0}. 
    \label{eq:RTE_spherical}
\end{equation}
The source function $j_\nu$ is defined as 
\begin{equation}
    j_\nu = \frac{1}{4\pi}\frac{\delta(r-r_{\rm s})}{4\pi r_{\rm s}^2}  \delta(x - x_{\rm i}), 
    \label{eq:jnu_sph}
\end{equation}
where the source uniformly distributes on a sphere of radius $r_{\rm s}$ and emits isotropically. 
The source function satisfies normalisation $\int {\rm d}\Omega\, {\rm d}r\, {\rm d}x\, (4\pi r^2j_\nu) =1$ under our definition. 
With the boundary condition of equation~(\ref{eq:bc}), the closed-form solution under spherical geometry is 
\begin{equation}
\begin{split}
    J_{\rm sph} = \frac{1}{4\pi r_0^2}\frac{\sqrt{6}x^2}{12 \sqrt{\pi} a_{\rm v}\tau_0}
    \frac{\sin \left( \pi\frac{\tau_{\rm s}}{\tau_0} \right) \left/ \left( \frac{\tau_{\rm s}}{\tau_0} \right) \right. }
    {\cos\left( \frac{\pi\tau_{\rm s}}{\tau_0} \right) + \cosh \left( \sqrt{ \frac{2\pi^3}{27}} \frac{ \left|x^3 - x_{\rm i}^3 \right|}{a_{\rm v}\tau_0} \right)}. 
    \label{eq:sph_full}
\end{split}
\end{equation}
This solution extends the work of \cite{Dijkstra_2006} to the $x_{\rm i} \neq 0$ case. 

Comparing equation~(\ref{eq:cyl_full}), (\ref{eq:pla_full}), and (\ref{eq:sph_full}), we notice they share very similar functional forms. 
The differences mainly reside at two places. 
One is the first fraction of each solution, which is relevant to the gauge of the boundary for a given geometry. 
The factors $2$, $2\pi r_0$, and $4\pi r_0^2$ correspond to two sides of a plane-parallel slab, the circumference of a cylinder, and the surface area of a sphere, respectively. 
The other difference is in the numerator of the last fraction in each solution, which affects the width of  $\Lya$ spectra. 

Fig.~\ref{fig:spec_3param} shows $\Lya$ spectra under three geometries, by varying each of $\tau_0$, $\tau_{\rm s}$ and $x_{\rm i}$ at a time. 
Compared with the RT simulation results, the analytic solutions work well at a similar level for the three geometries.
For a given $\tau_0$, both the width of each peak and the separation of the two peaks are the largest for slab geometry and the smallest for spherical geometry. 
This is consistent with the functional forms, as well as our expectation. 
Under spherical geometry, the photons have three spatial dimensions to escape, compared to two under cylindrical geometry and one under slab geometry. 
As a result, for given $\tau_0$, $\Lya$ photons in a spherical cloud experience the least number of scatterings among three geometries before escaping, leading to the least diffusion in the frequency space. 

In the last column of Fig.~\ref{fig:spec_3param}, the shift of the initial frequency from the lab-frame line centre ($x_{\rm i} \neq 0$) leads to asymmetric line profiles. 
For a given $x_{\rm i}$, the ratio of flux at the higher peak to that at the lower peak is the smallest for the slab geometry (largest number of scatterings) and the largest for the spherical geometry (smallest number of scatterings), which reflects that scatterings tend to erase the initial asymmetry in frequency space. 
\section{The effect of atomic recoil}\label{sec:recoil}

\begin{figure}
    \includegraphics[alt={A graph showing the ratio of simulated Lyman-alpha spectra to those without from a cylindrical gas cloud for three optical depth values. The ratio can well characterised by an exponential function.}, width=\columnwidth]{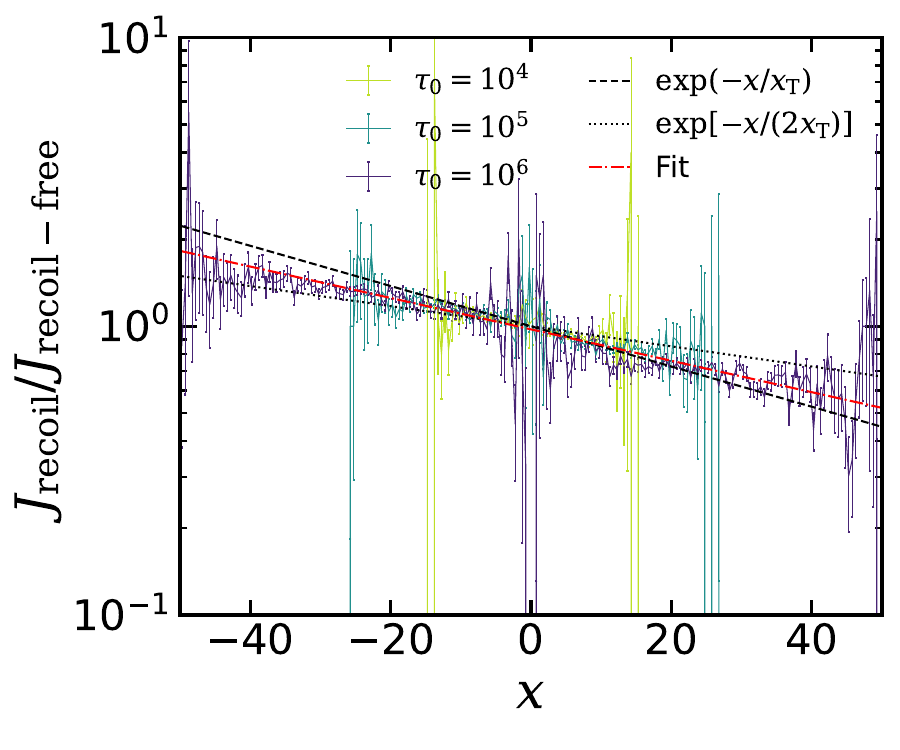}
    \caption{Ratio $J_{\rm recoil}/J_{\rm recoil-free}$ of simulated $\Lya$ spectra with  recoil to those without from a cylindrical gas cloud for three optical depth values. 
    The error bars are estimated based on the Poisson noise of $\Lya$ RT simulations in each frequency bin. 
    The dashed line and dotted line correspond to $\exp(-x/x_{\rm T})$ and $\exp{[-x/(2 x_{\rm T})]}$, respectively. 
    The ratio $J_{\rm recoil}/J_{\rm recoil-free}$ approximately lies between the two exponential functions. 
    The dash-dotted line is obtained by jointly fitting an exponential function $D \exp(-\alpha x/x_{\rm T})$ to the three coloured lines with different $\tau_0$, which yields $D=0.98$ and $\alpha=0.78$. 
    }
    \label{fig:cyl_RecNrec_ratio}
\end{figure}

\begin{figure*}
    \includegraphics[alt={A graph showing comparisons between simulated Lyman-alpha spectra with recoil and those from modifying the recoil-free simulated Lyman-alpha spectra with an exponential function. The modified Lyman-alpha spectra agree well with simulated ones for various optical depths, source positions, initial frequencies, and geometries.}, width=\textwidth]{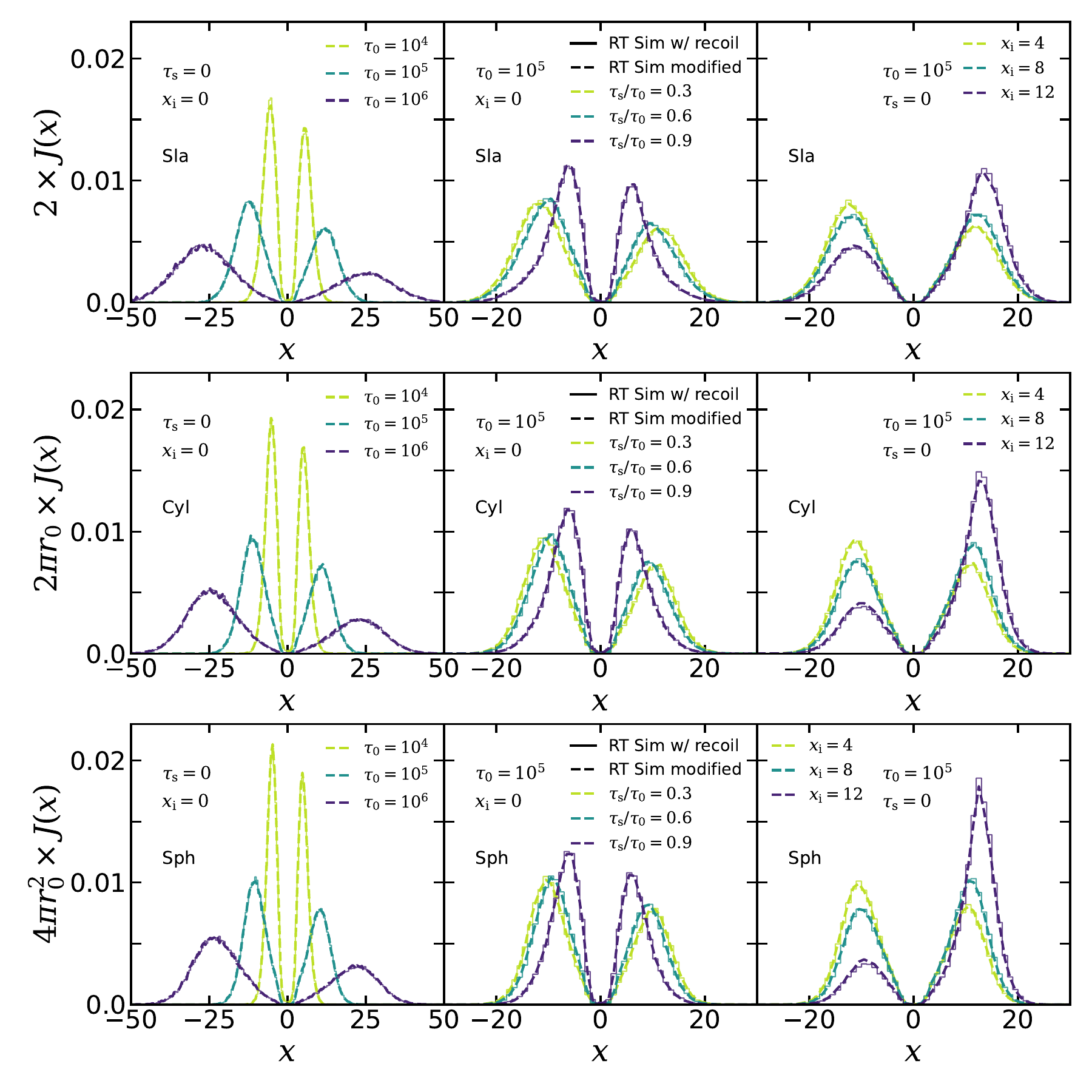}
    \caption{Comparisons between simulated $\Lya$ spectra with recoil and those from modifying the recoil-free simulated $\Lya$ spectra with an exponential function (see Section~\ref{sec:recoil}) for various optical depths $\tau_0$, source positions $\tau_{\rm s}$, initial frequencies $x_{\rm i}$, and geometries. 
    The three rows correspond to slab, cylindrical, and spherical geometries, respectively. 
    The three columns correspond to varying one of the three parameters, $\tau_0$, $\tau_{\rm s}/\tau_0$, and $x_{\rm i}$, respectively, with the rest two parameters fixed as labelled in each panel. 
    }
    \label{fig:spec_recoil_simcorr}
\end{figure*}

Solving $\Lya$ RT equations with atomic recoil can be difficult \citep[e.g.][]{Nebrin_2025}. 
In this section, we propose to account for recoil effect for a static, uniform cloud under different geometries by multiplying recoil-free $\Lya$ RT spectra with a modification term. 
We first obtain the functional form of the modification term by analysing the mathematical structure of the $\Lya$ RT equation with recoil, then demonstrate that the functional form is consistent with results from $\Lya$ RT simulations. 
Finally, we apply the modification to recoil-free $\Lya$ spectra from $\Lya$ RT simulations under various optical depths, source positions, initial frequencies, and geometries, and compare them to simulated $\Lya$ spectra with recoil. 

We start from the $\Lya$ RT equation with recoil under cylindrical geometry following \cite{Nebrin_2025},
\begin{equation}
    \frac{\partial^2 J}{\partial \tau^2} + \frac{1}{\tau}\frac{\partial J}{\partial \tau} + \frac{\partial^2 J}{\partial y^2} + \frac{\sqrt{6}}{2x_{\rm T}} \frac{\partial (\voigt J)}{\partial y}= -\frac{3\voigt j_\nu}{n_0\sigma_0}, 
    \label{eq:RTE_recoil}
\end{equation}
where $x_{\rm T} \equiv (k_{\rm B}T)/(h\Delta\nu_{\rm D})$ \citep{Zheng_2002a}\footnote{The notation $x_{\rm T}$ corresponds to $\bar{x}$ in \cite{Nebrin_2025}, with $\bar{x} = 1/(2 x_{\rm T})$.}.
The recoil effect is described by the term containing $x_{\rm T}$.
We first separate variables following $J(y,\tau) = \sum_{n=0}^{\infty} E_n(\tau) f_n(y)$. 
The recoil term does not affect the procedure for obtaining $E_n(\tau)$, while the equation for $f_n(y)$ becomes 
\begin{equation}
    \frac{{\rm d}^2 f_n}{{\rm d} y^2} + \frac{\sqrt{6}}{2 x_{\rm T}}\frac{{\rm d}(\voigt f_n)}{{\rm d}y}- \lambda_n^2 f_n = -\frac{3\voigt Q_n}{n_0 \sigma_0}. 
    \label{eq:ses_fn_recoil}
\end{equation}
To reveal the hidden structure in equation~(\ref{eq:ses_fn_recoil}), we adopt the following trial solution $f_n (y)= h_n(y) \mathcal{E} (y)$, where we require $h_n(y)$ to be the recoil-free solution and satisfy equation~(\ref{eq:ses_hn}). 
With this trial function, equation~(\ref{eq:ses_fn_recoil}) becomes 
\begin{equation}
\begin{split}
    &\mathcal{E} \frac{{\rm d}^2 h_n}{{\rm d} y^2} 
    + \left( 2\frac{{\rm d}\mathcal{E}}{{\rm d}y} + \frac{\sqrt{6}}{2 x_{\rm T} } \voigt \mathcal{E} \right) \frac{{\rm d}h_n}{{\rm d}y} \\
    &+ \left[ \frac{{\rm d}^2\mathcal{E}}{{\rm d} y^2} + \frac{\sqrt{6}}{2 x_{\rm T} } \frac{{\rm d} (\voigt \mathcal{E})}{{\rm d} y}  - \lambda_n^2 \mathcal{E} \right] h_n 
    = -\frac{3\voigt Q_n}{n_0 \sigma_0}. 
\end{split}
    \label{eq:ses_hnE_recoil}
\end{equation}

Without explicitly solving equation~(\ref{eq:ses_hnE_recoil}), we derive an approximate functional form for $\mathcal{E}$ with the following procedure. 
First, we evaluate both equation~(\ref{eq:ses_hnE_recoil}) and (\ref{eq:ses_hn}) at $y \neq y_{\rm i}$, which leads to $Q_n = 0$. 
Second, we multiply equation~(\ref{eq:ses_hn}) by $\mathcal{E}$ and subtract equation~(\ref{eq:ses_hnE_recoil}). 
Third, we switch the frequency variable $y$ back to $x$. 
These steps lead to 
\begin{equation}
    2 \voigt \frac{{\rm d}\mathcal{E}}{{\rm d}x} \frac{{\rm d}h_n}{{\rm d}x}  + h_n \frac{{\rm d}}{{\rm d}x}\left( \voigt \frac{{\rm d}\mathcal{E}}{{\rm d}x} \right) + \frac{1}{x_{\rm T}}\frac{{\rm d} (\voigt h_n \mathcal{E})}{{\rm d}x} = 0.
    \label{eq:Ex}
\end{equation}
In one special case where the first term is much smaller than the second term in equation~(\ref{eq:Ex}), we can modify the first two terms in equation~(\ref{eq:Ex}) as  
\begin{equation}
\begin{split}
    & 2 \voigt \frac{{\rm d}\mathcal{E}}{{\rm d}x} \frac{{\rm d}h_n}{{\rm d}x}  + h_n \frac{{\rm d}}{{\rm d}x}\left( \voigt \frac{{\rm d}\mathcal{E}}{{\rm d}x} \right) \\
    & \simeq \voigt \frac{{\rm d}\mathcal{E}}{{\rm d}x} \frac{{\rm d}h_n}{{\rm d}x}  + h_n \frac{{\rm d}}{{\rm d}x}\left( \voigt \frac{{\rm d}\mathcal{E}}{{\rm d}x} \right) 
    = \frac{{\rm d}}{{\rm d}x}\left( h_n \voigt \frac{{\rm d}\mathcal{E}}{{\rm d}x} \right).
\end{split}
    \label{eq:Ex_approx1}
\end{equation}
Plugging equation~(\ref{eq:Ex_approx1}) back to equation~(\ref{eq:Ex}) yields 
\begin{equation}
    \frac{{\rm d}}{{\rm d}x}\left[ h_n \voigt \left( \frac{{\rm d}\mathcal{E}}{{\rm d}x} + \frac{1}{x_{\rm T}} \mathcal{E} \right) \right] = 0. 
    \label{eq:Ex_approx_eq1}
\end{equation}
The functional form of $\mathcal{E}$ can be obtained from equation~(\ref{eq:Ex_approx_eq1}) as $\mathcal{E} \propto \exp(-x/x_{\rm T})$\footnote{
Equation~(\ref{eq:Ex_approx_eq1}) leads to $h_n \voigt ({\rm d}\mathcal{E}/{\rm d}x + \mathcal{E}/x_{\rm T}) = \rm constant$. 
We argue that the constant has to be zero given both $\voigt$ and $h_n\mathcal{E}$ are expected to be zero as $|x| \to \infty$, which leads to ${\rm d}\mathcal{E}/{\rm d}x + \mathcal{E}/x_{\rm T}=0$. 
}.

The factor $\mathcal{E} \propto \exp(-x/x_{\rm T})$ is not unique. 
In another special case where the second term is much smaller than the first term in equation~(\ref{eq:Ex}), we can modify the first two terms in equation~(\ref{eq:Ex}) in a different way 
\begin{equation}
\begin{split}
    & 2 \voigt \frac{{\rm d}\mathcal{E}}{{\rm d}x} \frac{{\rm d}h_n}{{\rm d}x}  + h_n \frac{{\rm d}}{{\rm d}x}\left( \voigt \frac{{\rm d}\mathcal{E}}{{\rm d}x} \right) \\
    & \simeq 2\voigt \frac{{\rm d}\mathcal{E}}{{\rm d}x} \frac{{\rm d}h_n}{{\rm d}x}  + 2h_n \frac{{\rm d}}{{\rm d}x}\left( \voigt \frac{{\rm d}\mathcal{E}}{{\rm d}x} \right) 
    = \frac{{\rm d}}{{\rm d}x}\left( 2h_n \voigt \frac{{\rm d}\mathcal{E}}{{\rm d}x} \right).
\end{split}
    \label{eq:Ex_approx2}
\end{equation}
Following similar steps of equation~(\ref{eq:Ex_approx1})--(\ref{eq:Ex_approx_eq1}), we obtain $\mathcal{E} \propto \exp[-x/(2 x_{\rm T})]$ based on the assumption made in equation~(\ref{eq:Ex_approx2}). 
From the analysis above, even though we do not explicitly solve $\mathcal{E}$ from equation~(\ref{eq:ses_hnE_recoil}), we argue that the functional form should closely follow an exponential function $\exp(-\alpha x/x_{\rm T})$, where $\alpha \sim 1$. 
This result is consistent with the physical picture that the scattering tends to thermalise $\Lya$ photons to follow $\HI$ gas temperature as argued by \cite{Zheng_2002a}, who suggest the form of $\exp(-x/x_{\rm T})$. 
This form can also be obtained following \cite{Nebrin_2025}, who solve the $\Lya$ RT equation with recoil under spherical geometry in the wing limit $|x| \gg 1$. 
By keeping the first three leading-order terms of $x$ in their solution (see their equation (A25) and (A36) as $|z| \ll 1$), the parameter $\alpha=1$ can be naturally derived.

To verify the insights obtained above, we run $\Lya$ RT simulations with and without recoil for a static, uniform gas cloud under cylindrical geometry. 
As shown in Fig.~\ref{fig:cyl_RecNrec_ratio}, we compute the ratio of $\Lya$ spectra with recoil to those without recoil for various optical depths and fixed $\tau_{\rm s} = 0$ and $x_{\rm i}=0$. 
The ratio values for three optical depths share the same trend on the plot, which is between $\exp(-x/x_{\rm T})$ (dashed line) and $\exp[-x/(2x_{\rm T})]$ (dotted line). 
The slope between the logarithmic ratio and frequency $x$ varies between small and large $|x|$, and approaches the trend of $\exp[-x/(2x_{\rm T})]$ towards large $|x|$. 
The variation of the slope, however, is not significant. 
Therefore, we choose to apply a single exponential function $D\exp(-\alpha x/x_{\rm T})$ to characterise the ratio. 
The free parameters $D$ and $\alpha$ are obtained by jointly fitting this function to the three lines from $\Lya$ RT simulations on Fig.~\ref{fig:cyl_RecNrec_ratio}, which results in $D=0.98$ and $\alpha=0.78$. 
The best-fit function is plotted as the dash-dotted line, which agrees well with $\Lya$ RT simulations across a large range of $x$.

Based on the exercise above, we account for the recoil effect in $\Lya$ spectra $J_{\rm cyl, r}$ by modifying recoil-free $\Lya$ spectra $J_{\rm cyl}$ following $J_{\rm cyl,r}=D_{\rm cyl}J_{\rm cyl} \exp(-\alpha x/x_{\rm T})$. 
The factor $D_{\rm cyl}$ is determined by requiring the same normalisation between $J_{\rm cyl, r}$ and $J_{\rm cyl}$ to reflect photon number conservation, such that $\int_{-\infty}^{+\infty} {\rm d}x J_{\rm cyl}= \int_{-\infty}^{+\infty} {\rm d}x D_{\rm cyl} J_{\rm cyl} \exp{(-\alpha x/x_{\rm T})}$. 
With such a procedure, the factor $D_{\rm cyl}$ is derived through $\left( \int_{-\infty}^{+\infty} {\rm d}x J_{\rm cyl} \right) \left/ \left[\int_{-\infty}^{+\infty} {\rm d}x J_{\rm cyl} \exp(-\alpha x/x_{\rm T}) \right] \right.$, depending on $\tau_0$, $\tau_{\rm s}$ and $x_{\rm i}$. 
The above procedure can be executed similarly for the slab and spherical geometries, which will introduce normalisation factors $D_{\rm sla}$ and $D_{\rm sph}$, respectively. 
The parameter $\alpha$ is fixed at 0.78 for all three geometries. 

To examine the performance of the modification method, we apply the modification to recoil-free spectra from $\Lya$ RT simulations under various optical depths, source positions, initial frequencies, and geometries. 
Then we compare the $\Lya$ spectra after modification to simulated $\Lya$ spectra with recoil, as shown in Fig.~\ref{fig:spec_recoil_simcorr}. 
The $\Lya$ spectra after modification agree well with $\Lya$ RT simulations for most cases, with only minor deviations near the peaks for the largest $x_{\rm i}$ in consideration. 
\section{The effect of velocity gradient}\label{sec:vgrad}
\begin{figure*}
    \includegraphics[alt={A graph showing Lyman-alpha spectra under non-zero velocity gradients evaluated by radiative transfer simulations, fitting formulae, and series solutions for various optical depths, geometries, and velocity gradients. The fitting formulae agree well with simulations across a broad range of velocity gradients. The series solutions agree well with radiative transfer simulations for small velocity gradients but show deviations for large ones.}, width=\textwidth]{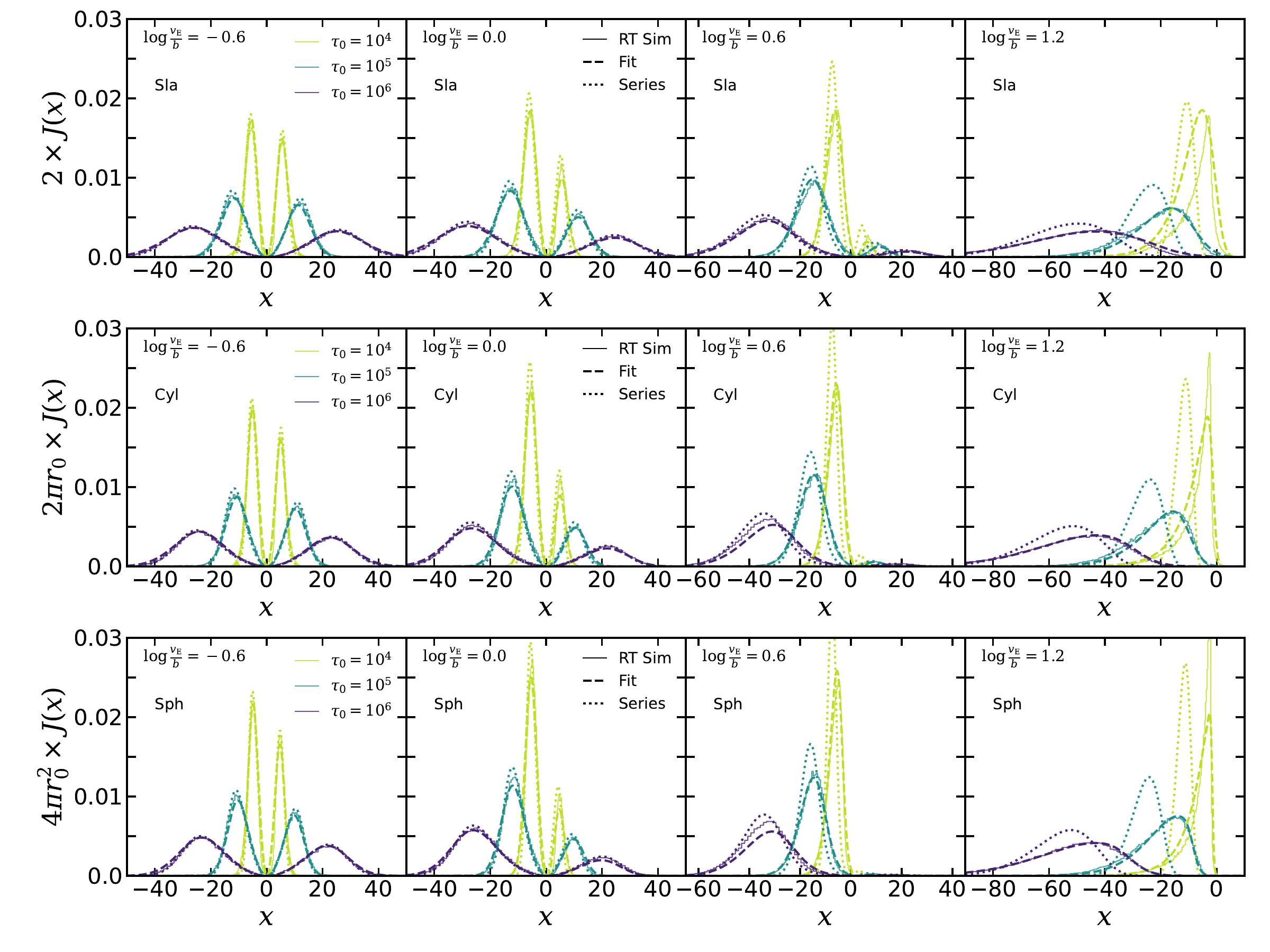}
    \caption{$\Lya$ spectra under constant velocity gradients for various optical depths $\tau_0$ (different colours) and geometries (different rows). 
    The source position $\tau_{\rm s}$ and initial frequency $x_{\rm i}$ are both fixed at 0. 
    The solid, dashed, and dotted lines correspond to $\Lya$ spectra from $\Lya$ RT simulations, fitting formulae, equation~(\ref{eq:ana_central_ext_pla}), (\ref{eq:ana_central_ext}), and (\ref{eq:ana_central_ext_sph}), and series solutions, equation~(\ref{eq:sesv_full_sla}), (\ref{eq:sesv_full}), and (\ref{eq:sesv_full_sph}), respectively. From left to right columns, the velocity gradient increases, as indicated by the ratio of velocity at the cloud edge to the thermal velocity, $v_{\rm E}/b$.}
    \label{fig:spec_tau0_vratio}
\end{figure*}

\begin{figure*}
    \includegraphics[alt={A graph showing Lyman-alpha spectra under non-zero velocity gradients evaluated by radiative transfer simulation and series solutions for various optical depths, source positions, initial frequencies, geometries, and velocity gradients. The series solutions agree well with radiative transfer simulations for small velocity gradients but show deviations for large ones.}, width=\textwidth]{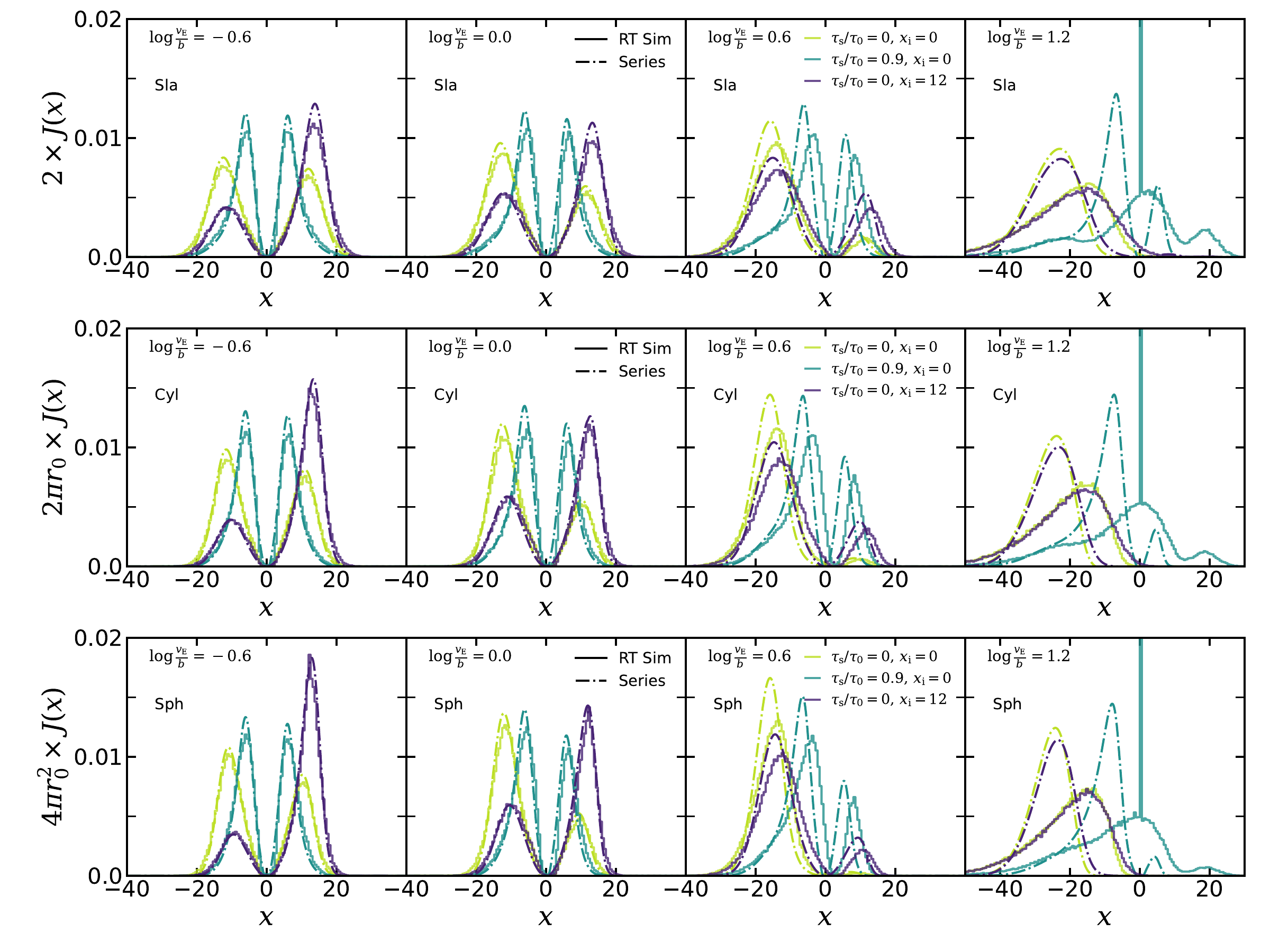}
    \caption{$\Lya$ spectra under constant velocity gradients for varying source positions $\tau_{\rm s}$ and initial frequencies $x_{\rm i}$ as labelled by the colours. 
    The optical depth is fixed at $\tau_0=10^5$. 
    The solid lines correspond to $\Lya$ spectra from $\Lya$ RT simulations, while the dash-dotted lines are from series solutions, equation~(\ref{eq:sesv_full_sla}), (\ref{eq:sesv_full}), and (\ref{eq:sesv_full_sph}). 
    The three rows correspond to slab, cylindrical, and spherical geometries, respectively. From left to right columns, the velocity gradient increases, as indicated by the ratio of velocity at the cloud edge to the thermal velocity, $v_{\rm E}/b$.
    }
    \label{fig:spec_vratio_taus_xi}
\end{figure*}

Gas bulk motion leaves strong imprints on $\Lya$ spectra. 
In this section, we solve $\Lya$ RT equations for a uniform gas cloud with a constant velocity gradient under different geometries. 
The solutions work well for small velocity gradients but fail at the large end. 
To characterise $\Lya$ spectra under large velocity gradients, we empirically extend the solutions by fitting simulated $\Lya$ spectra, which perform reasonably well across a large range of velocity gradients and optical depths and show significant improvement for large velocity gradients under large optical depths. 

\subsection{Solutions under constant velocity gradients}\label{subsec:ses_vgrad}

We start from the $\Lya$ RT equation under cylindrical geometry closely following \cite{Nebrin_2025}, 
\begin{equation}
    \frac{\partial^2 J}{\partial \tau^2} + \frac{1}{\tau}\frac{\partial J}{\partial \tau} + \frac{\partial^2 J}{\partial y^2} + \gamma \frac{\partial J}{\partial y}= -\frac{3\voigt j_\nu}{n_0\sigma_0}, 
    \label{eq:RTEV_cylindrical}
\end{equation}
where the constant velocity gradient introduces an additional term $\gamma (\partial J/\partial y)$, with $\gamma \equiv \sqrt{6}(\nabla \cdot \bm{u})/(3b n_0 \sigma_0) = (2\sqrt{6}v_{\rm E})/(3b \tau_0)$. 
The vector $\bm{u}$ is the three-dimensional velocity field, which satisfies $\nabla \cdot \bm{u}=2v_{\rm E}/r_0$ under cylindrical geometry. 
The parameter $v_{\rm E}$ is the gas velocity at the edge of the cloud along the radial direction of the cylindrical cloud, with positive $v_{\rm E}$ representing expanding clouds and negative $v_{\rm E}$ representing contracting clouds. 
The source function $j_\nu$ has the same definition as in equation~(\ref{eq:jnu}). 
Equation~(\ref{eq:RTEV_cylindrical}) only applies when $|v_{\rm E}/b| \ll (a_{\rm v}\tau_0)^{2/3}/12$ as discussed in \cite{Nebrin_2025}. 

The solving procedure to equation~(\ref{eq:RTEV_cylindrical}) with velocity gradients closely resembles that to equation~(\ref{eq:RTE_cylindrical}) without velocity gradients. 
For brevity, we only describe the key steps as follows. 
The first step is still separating variables as shown by equation~(\ref{eq:ses_expansion}), followed by identical procedures to obtain the same $E_n(\tau)$ in equation~(\ref{eq:En_general}). 
The frequency function $h_n(y)$, however, satisfies the following new equation 
\begin{equation}
    \frac{{\rm d}^2 h_n}{{\rm d} y^2} +\gamma \frac{{\rm d} h_n}{{\rm d} y} - \lambda_n^2 h_n = -\frac{3\voigt Q_n}{n_0 \sigma_0}. 
    \label{eq:sesv_hn}
\end{equation}
The solution is 
\begin{equation}
    h_n(y) = \frac{\sqrt{3}n_0 \sigma_0}{8\pi^2 \tau_0 \gamma_n} \frac{J_0(\lambda_n \tau_{\rm s})}{|J_1(\lambda_n \tau_0)|}
    \exp \left[-\frac{\gamma}{2}(y-y_{\rm i}) -\gamma_n |y - y_{\rm i}| \right], 
    \label{eq:hnv_general}
\end{equation}
where $\gamma_n \equiv \sqrt{\lambda_n^2+\gamma^2/4}$ with $\lambda_n=(n+3/4)\pi/\tau_0$. 
% The coefficient $C^\prime$ can determined by replacing $C$ with $C^\prime$ and $\lambda_n$ with $\gamma_n$ in equation~(\ref{eq:jump}). 
Finally, by combining $E_n(\tau)$ in equation~(\ref{eqn:En}) and $h_n(y)$ in equation~(\ref{eq:hnv_general}), followed by evaluating the function at the boundary of the cloud, $\tau=\tau_0$, the series solution for the emergent spectrum in a uniform cloud with a constant velocity gradient under cylindrical geometry is found as 
\begin{equation}
\begin{split}
    J_{\rm cyl, v}(y) = &\frac{1}{2\pi r_0}\frac{\sqrt{6} x^2}{6\sqrt{\pi} a_{\rm v}\tau_0} \exp \left[-\frac{\gamma}{2}(y-y_{\rm i}) \right] \\
    & \sum_{n=0}^{\infty}\frac{J_0(\lambda_n \tau_{\rm s})}{J_1(\lambda_n \tau_0)} \frac{\lambda_n}{\gamma_n} \exp(-\gamma_n |y - y_{\rm i}|). 
    \label{eq:sesv_full}
\end{split}
\end{equation}

Comparing equation~(\ref{eq:sesv_full}) to equation~(\ref{eq:ses_full}), the constant velocity gradient introduces $\exp \left[-\gamma(y-y_{\rm i}) /2\right]$ governing the flux ratio of the two peaks in $\Lya$ spectra, with an expanding cloud (positive $\gamma$) suppressing the blue peak (positive $y$) while a contracting cloud suppressing the red peak (negative $y$). 
The limiting behaviour of equation~(\ref{eq:sesv_full}) near zero velocity gradient is also interesting. 
When $|\gamma| \ll 2\lambda_n$ (or $|v_{\rm E}| \ll 3b$), the coefficient $\gamma_n$ approaches $\lambda_n$. 
As a consequence, the solution with a small velocity gradient can be viewed as a simple modification to that without velocity gradients, which satisfies $J_{\rm cyl, v} \approx J_{\rm cyl} \exp \left[-\gamma(y-y_{\rm i})/2 \right]$. 

Similar to the case of cylindrical geometry, the solutions can also be obtained for $\Lya$ spectra from a uniform cloud with a constant velocity gradient under slab and spherical geometries. 
We list the key steps here. 
For slab geometry, the $\Lya$ RT equation is 
\begin{equation}
    \frac{\partial^2 J}{\partial \tau^2} + \frac{\partial^2 J}{\partial y^2} + \gamma \frac{\partial J}{\partial y}= -\frac{3\voigt j_\nu}{n_0\sigma_0}, 
    \label{eq:RTEV_slab}
\end{equation}
with $j_\nu$ the same as in equation~(\ref{eq:jnu_pla}) and $\gamma \equiv\sqrt{6}(\nabla \cdot \bm{u})/(3bn_0 \sigma_0)$ sharing the same definition as that for cylindrical geometry. 
However, due to the difference in geometry, the velocity vector $\bm{u}$ only has one non-zero component along the finite dimension of the slab, which leads to a change in the expression of $\gamma = (\sqrt{6}v_{\rm E})/(3b\tau_0)$. 
The edge velocity $v_{\rm E}$ is measured along the finite dimension, with positive values at $r=r_0$ representing an expanding slab while negative values at $r=r_0$ representing a contracting slab. 
The solution to equation~(\ref{eq:RTEV_slab}) is 
\begin{equation}
\begin{split}
    J_{\rm sla, v}(y) = &\frac{1}{2}\frac{\sqrt{6}x^2}{6 \sqrt{\pi} a_{\rm v}\tau_0} \exp \left[-\frac{\gamma}{2}(y-y_{\rm i}) \right] \\
    \times & \sum_{n=0}^{\infty} (-1)^n \cos(\lambda_n \tau_{\rm s}) \frac{\lambda_n}{\gamma_n} \exp(-\gamma_n |y - y_{\rm i}|), 
    \label{eq:sesv_full_sla}
\end{split}
\end{equation}
with $\lambda_n=(n+1/2)\pi/\tau_0$ \citep[e.g.][]{Harrington_1973} and $\gamma_n \equiv \sqrt{\lambda_n^2+\gamma^2/4}$ updated according to $\lambda_n$ and $\gamma$. 

For spherical geometry, the solution is discussed in depth in \cite{Nebrin_2025} and \cite{Smith_2025}. 
We present the key steps for convenient comparisons to other geometries. 
The $\Lya$ RT equation is 
\begin{equation}
    \frac{\partial^2 J}{\partial \tau^2} + \frac{2}{\tau}\frac{\partial J}{\partial \tau} + \frac{\partial^2 J}{\partial y^2} + \gamma \frac{\partial J}{\partial y}= -\frac{3\voigt j_\nu}{n_0\sigma_0}, 
    \label{eq:RTEV_sph}
\end{equation}
with $j_\nu$ the same as in equation~(\ref{eq:jnu_sph}) and $\gamma \equiv\sqrt{6}(\nabla \cdot \bm{u})/(3bn_0 \sigma_0)$. 
Evaluating $\nabla \cdot \bm{u}=3v_{\rm E}/r_0$ leads to $\gamma = (\sqrt{6}v_{\rm E})/(b\tau_0)$, with the edge velocity $v_{\rm E}$ measured along the radial direction of the sphere. 
The solution to equation~(\ref{eq:RTEV_sph}) is 
\begin{equation}
\begin{split}
    J_{\rm sph, v}(y) = &\frac{1}{4\pi r_0^2}\frac{\sqrt{6}x^2}{6 \sqrt{\pi} a_{\rm v}\tau_0} \exp \left[-\frac{\gamma}{2}(y-y_{\rm i}) \right] \\
    \times & \sum_{n=0}^{\infty} (-1)^n\frac{\sin(\lambda_n \tau_{\rm s})}{\tau_{\rm s}/\tau_0} \frac{\lambda_n}{\gamma_n} \exp(-\gamma_n |y - y_{\rm i}|), 
    \label{eq:sesv_full_sph}
\end{split}
\end{equation}
with $\lambda_n=(n+1)\pi/\tau_0$ \citep{Dijkstra_2006, Nebrin_2025} and $\gamma_n \equiv \sqrt{\lambda_n^2+\gamma^2/4}$ updated according to $\lambda_n$ and $\gamma$. 

Fig.~\ref{fig:spec_tau0_vratio} compares $\Lya$ spectra from equation~(\ref{eq:sesv_full_sla}), (\ref{eq:sesv_full}), and (\ref{eq:sesv_full_sph}) with dotted lines, to those from RT simulations with solid lines, for different $\tau_0$ and $v_{\rm E}/b$, fixing $\tau_{\rm s}$ and $y_{\rm i}$ at 0. 
They agree well at small velocity gradients, but loses accuracy as velocity gradients become larger. 
There are multiple features emergent on $\Lya$ spectra under large velocity gradients, including strong flux suppression of one peak, extended tails at large $|x|$, and non-monotonic dependence of the peak position on velocity gradient. 
To characterise $\Lya$ spectra under large velocity gradients, it requires analytical formulae that can properly address these emergent features. 

Equation~(\ref{eq:sesv_full_sla}), (\ref{eq:sesv_full}), and (\ref{eq:sesv_full_sph}) also extend the solutions in literature and describe the dependence of $\Lya$ spectra on source position $\tau_{\rm s}$ and initial frequency $x_{\rm i}$. 
Fig.~\ref{fig:spec_vratio_taus_xi} compares these solutions to $\Lya$ RT simulations for different values of $\tau_{\rm s}$ and $x_{\rm i}$ for each geometry. 
Similar to the discussions of Fig.~\ref{fig:spec_tau0_vratio}, the solutions work well under small velocity gradients and show deviations under large ones. 
The features of $\Lya$ spectra (e.g. the peak position, the shape of peaks, and the flux ratio of two peaks) show complicated dependence on velocity gradients, source positions, and initial frequencies, indicating the importance of understanding $\Lya$ RT when interpreting observed $\Lya$ spectra. 
We further notice the spike feature near $x=0$ for spectra from the $\Lya$ RT simulations with $\tau_{\rm s}/\tau_0=0.9$ in the last column of Fig.~\ref{fig:spec_vratio_taus_xi}. 
For those cases, the photons emitted outwards can easily escape the gas cloud given that the source position is close to the edge of the cloud. 
Moreover, the initial frequency of $\Lya$ photons is at the line centre in the lab frame, which shifts away from the line centre in the rest frame of the $\HI$ gas due to strong outflow. 
Both effects lead to almost direct escaping of a fraction of $\Lya$ photons after being emitted and create a spike feature on spectra near the line centre of the lab frame.  The series solutions are not able to capture such features, as the condition of dominance by wing scatterings (extremely optically thick) used in deriving the corresponding $\Lya$ RT equations no longer holds.

\subsection{Fitting formulae under constant velocity gradients}\label{subsec:fitting_vgrad}
To describe $\Lya$ spectra under large velocity gradients, we extend the functional forms of the closed-form solution under small velocity gradients, limiting to $\tau_{\rm s}=0$ and $y_{\rm i}=0$. 
We first use the cylindrical case as an example to demonstrate steps for obtaining such functional forms. 
Then we provide the fitting formulae for obtaining $\Lya$ spectra through the functional forms. The results for the slab and spherical cases are also presented following similar procedures.  

As discussed in Section~\ref{subsec:ses_vgrad}, $\Lya$ spectra under small velocity gradient can be described by multiplying an exponential factor $\exp[-\gamma(y-y_{\rm i})/2]$ to those without velocity gradient (equation~(\ref{eq:cyl_full})). 
With $\tau_{\rm s}=0$ and $y_{\rm i}=0$, this leads to 
\begin{equation}
\begin{split}
                J_{\rm cyl, v} \approx &\frac{1}{2 \pi r_0}\frac{\sqrt{6}x^2}{12 \sqrt{\pi} a_{\rm v}\tau_0} 
                \frac{\exp{ \left(-\frac{2\sqrt{\pi}}{9} \frac{v_{\rm E}}{b} \frac{ x^3}{a_{\rm v}\tau_0} \right)}}
                {1 + \cosh \left( \sqrt{ \frac{2\pi^3}{27}} \frac{ \left|x^3\right|}{a_{\rm v}\tau_0} \right)} \\
                \times &\left\{ 1.92\exp{ \left(\frac{1}{4} \sqrt{ \frac{2\pi^3}{27}} \frac{ \left|x^3\right|}{a_{\rm v}\tau_0} \right)} \right. \\
                       & + 1.32\exp{ \left(-\frac{3}{4} \sqrt{ \frac{2\pi^3}{27}} \frac{ \left|x^3\right|}{a_{\rm v}\tau_0} \right)} \\
                       &\left. -0.609\exp{ \left(-\frac{7}{4} \sqrt{ \frac{2\pi^3}{27}} \frac{ \left|x^3\right|}{a_{\rm v}\tau_0} \right)} \right\}. 
            \end{split}
    \label{eq:ana_central}
\end{equation}
Now, we extend this formula with the following steps. 
First, we parametrize all $|x^3|$ term to be $c_{\lambda}|x|^{n_{\lambda}}$, which controls how intensity $J$ decreases with frequency $x$ at the tail of a $\Lya$ spectrum (i.e. $|x| \gg 1$). 
Second, we notice the divergent behaviour caused by $\exp{ \{-(2\sqrt{\pi}/9) (v_{\rm E}/b)  [x^3/(a_{\rm v}\tau_0) ]}\}$ when $|v_{\rm E}/b|$ is large. 
To address this problem, we replace it %$\exp{ \{-(2\sqrt{\pi}/9) (v_{\rm E}/b)  [x^3/(a_{\rm v}\tau_0) ]}\}$ 
with $1+{\rm erf}\{-(2\sqrt{\pi}/9) (v_{\rm E}/b) [x^3/(a_{\rm v}\tau_0)] \}$ utilizing the error function ${\rm erf}(z)$. 
Third, we parametrize $x^3$ inside the error function to be $(x/x_\gamma)^3/|x/x_\gamma|^{n_\gamma}$, and add a $c_\gamma$ to that term. 
This allows more flexibility to capture the change of intensity near the line centre $x=0$. 
Fourth, we change the $x^2$ term in equation~(\ref{eq:ana_central}) to be $[x - v_{\rm E}/(2b)]^2$ to address the slight shift of the frequency with zero intensity. 
Finally, we multiply equation~(\ref{eq:ana_central}) with a global normalisation factor $J_{\rm A}$ to match the amplitude. 
With these modifications, the formula after extension becomes 
\begin{equation}
\begin{split}
    J_{\rm cyl, vfit} = & \frac{J_{\rm A}}{2 \pi r_0} \frac{\sqrt{6}[x-v_{\rm E}/(2b)]^2}{12 \sqrt{\pi} a_{\rm v}\tau_0} \\
    \times & \frac{1 + {\rm erf}{ \left[ -\frac{2\sqrt{\pi}}{9} \frac{v_{\rm E}}{b} \frac{1}{a_{\rm v}\tau_0} \frac{(x/x_\gamma)^3}{|x/x_\gamma|^{n_\gamma}} +c_\gamma \right]}} 
                {1 + \cosh \left( \sqrt{ \frac{2\pi^3}{27}} \frac{ c_\lambda \left|x\right|^{n_\lambda}}{a_{\rm v}\tau_0} \right)} \\
                \times & \left\{ 1.92\exp{ \left(\frac{1}{4} \sqrt{ \frac{2\pi^3}{27}} \frac{ c_\lambda \left|x\right|^{n_\lambda}}{a_{\rm v}\tau_0} \right)} \right. \\
                    & + 1.32\exp{ \left(-\frac{3}{4} \sqrt{ \frac{2\pi^3}{27}} \frac{ c_\lambda\left|x\right|^{n_\lambda}}{a_{\rm v}\tau_0} \right)}  \\
                    &\left. -0.609\exp{ \left(-\frac{7}{4} \sqrt{ \frac{2\pi^3}{27}} \frac{ c_\lambda \left|x\right|^{n_\lambda}}{a_{\rm v}\tau_0} \right)} \right\}. 
            \end{split}
    \label{eq:ana_central_ext}
\end{equation}
In summary, there are six parameters applied to extend the analytical formula from low velocity gradients to high velocity gradients, namely $J_{\rm A}$, $c_\lambda$, $n_\lambda$, $x_\gamma$, $n_\gamma$, and $c_\gamma$, which jointly depend on $v_{\rm E}$ and $\tau_0$.  
Their values at low velocity gradients ($v_{\rm E}\to 0$) are well informed by equation~(\ref{eq:ana_central}), with $J_{\rm A} \sim 1$, $c_\lambda \sim 1$, $n_\lambda \sim 3$, $x_\gamma \sim (2/\sqrt{\pi})^{1/3} \sim 1$, $n_\gamma \sim 0$, and $c_\gamma \sim 0$. 

The extended functional form for slab and spherical geometries can be obtained similarly following the case of cylindrical geometry. 
For slab geometry, the formula becomes 
\begin{equation}
\begin{split}
    J_{\rm sla, vfit} = & \frac{J_{\rm A}}{2} \frac{\sqrt{6}[x-v_{\rm E}/(2b)]^2}{12 \sqrt{\pi} a_{\rm v}\tau_0} \\
                \times & \frac{1 + {\rm erf}{ \left[ -\frac{\sqrt{\pi}}{9} \frac{v_{\rm E}}{b} \frac{1}{a_{\rm v}\tau_0} \frac{(x/x_\gamma)^3}{|x/x_\gamma|^{n_\gamma}} +c_\gamma \right]}}
                {1 + \cosh \left( \sqrt{ \frac{2\pi^3}{27}} \frac{ c_\lambda \left|x\right|^{n_\lambda}}{a_{\rm v}\tau_0} \right)} \\
                \times & 2\cosh \left( \frac{1}{2} \sqrt{ \frac{2\pi^3}{27}} \frac{ c_\lambda \left|x\right|^{n_\lambda}}{a_{\rm v}\tau_0} \right). 
            \end{split}
    \label{eq:ana_central_ext_pla}
\end{equation}
For spherical geometry, the formula yields 
\begin{equation}
    \begin{split}
    J_{\rm sph, vfit} = & \frac{J_{\rm A}}{4\pi r_0^2} \frac{\sqrt{6}[x-v_{\rm E}/(2b)]^2}{12 \sqrt{\pi} a_{\rm v}\tau_0} \\
              \times & \frac{1 + {\rm erf}{ \left[ -\frac{\sqrt{\pi}}{3} \frac{v_{\rm E}}{b} \frac{1}{a_{\rm v}\tau_0} \frac{(x/x_\gamma)^3}{|x/x_\gamma|^{n_\gamma}} +c_\gamma \right]}}
              {1 + \cosh \left( \sqrt{ \frac{2\pi^3}{27}} \frac{ c_\lambda \left|x\right|^{n_\lambda}}{a_{\rm v}\tau_0} \right)} \times \pi. 
    \end{split}
    \label{eq:ana_central_ext_sph}
\end{equation}

After obtaining the functional form of $\Lya$ spectra in equation~(\ref{eq:ana_central_ext}), (\ref{eq:ana_central_ext_pla}), and (\ref{eq:ana_central_ext_sph}), the next task is to find the relations between physical parameters, $\tau_0$ and $v_{\rm E}$, and the six free parameters, $J_{\rm A}$, $c_\lambda$, $n_\lambda$, $x_\gamma$, $n_\gamma$, and $c_\gamma$, separately for each geometry. 
The procedure of obtaining fitting formulae for such relations are described in Appendix~\ref{sec:vgrad_fitfunc}, after which we can analytically calculate $\Lya$ spectra for given optical depth, velocity gradient, and geometry. 
These fitting formulae are also expected to work for different temperature considering that the temperature dependence is explicitly accounted for by the thermal velocity $b$ and Voigt parameter $a_{\rm v}$. 
This is tested by running $\Lya$ RT simulations under $T=10^4\, \rm K$ with $v_{\rm E}/b$ and $a_{\rm v} \tau_0$ matching the values shown in Fig.~\ref{fig:spec_tau0_vratio}. 
The resulting simulated spectra as a function of $x$ agree well between cases of $T=10\, \rm K$ and $T=10^4\, \rm K$. 

Fig.~\ref{fig:spec_tau0_vratio} shows $\Lya$ spectra from the fitting formulae with dashed lines for different $\tau_0$ and $v_{\rm E}$. 
They work well for small velocity gradients, and perform better than series solutions (dotted lines) on reproducing simulated $\Lya$ spectra (solid lines) for large velocity gradients, although a discrepancy starts to appear for low optical depths and large velocity gradients, as expected. 
The accuracy of the fitting formulae is quantified for different $v_{\rm E}$ and $\tau_0$ in Fig.~\ref{fig:err_fitfunc} of Appendix~\ref{sec:vgrad_fitfunc}. 
\section{Summary and discussion}\label{sec:summary}
In this paper, we study analytical formulae for solutions to $\Lya$ RT equations, accounting for the effects of geometry, recoil, and velocity gradient. 
We summarize our main results below. 
\begin{itemize}
\item[1.] We introduce the $\Lya$ RT equation for a uniform static cloud under cylindrical geometry. 
The series solution to this equation is derived, which characterises the dependence of emergent $\Lya$ spectra on optical depth $\tau_0$, source position $\tau_{\rm s}$, and initial frequency $x_{\rm i}$. 
We obtain closed functional forms by using the asymptotic forms of Bessel functions of the first kind and introduce a correction term to match the series solution. 
We also modify or extend solutions to $\Lya$ RT equations under slab and spherical geometries based on literature, as a function of $\tau_0$, $\tau_{\rm s}$, and $x_{\rm i}$. 
The solutions under three geometries reveal similar functional forms, with the main difference in the peak width and peak separation.  
We also compare $\Lya$ analytical solutions to Monte Carlo RT simulations, which demonstrates good consistency for all three geometries. 
With the above work, we complete the set of $\Lya$ analytical solutions for a static, uniform cloud under plane-parallel, cylindrical, and spherical symmetry, for various optical depths, source positions, and initial frequencies. 

\item[2.] Motivated by the $\Lya$ RT equation with recoil, we notice that the ratio of $\Lya$ spectra with recoil to recoil-free $\Lya$ spectra can be described by an exponential function of frequency $x$, which reflects the thermalisation of $\Lya$ photons to follow $\HI$ gas temperature and is justified by $\Lya$ RT simulations. 
Based on this insight, we propose a modification procedure to analytically account for the effect of recoil in $\Lya$ RT. 
The $\Lya$ spectrum with recoil is obtained by multiplying the recoil-free $\Lya$ spectrum with a function proportional to $\exp(-0.78x/x_{\rm T})$ with $x_{\rm T} \equiv (k_{\rm B}T)/(h\Delta\nu_{\rm D})$ and by renormalising the spectrum for photon number conservation. 
This simple procedure is applied to recoil-free $\Lya$ spectra from $\Lya$ RT simulations, which demonstrates good agreement to those with recoil under various optical depths, source postions, initial frequencies, and geometries.

\item[3.] To study $\Lya$ spectra from gas with bulk motion, we obtain series solutions to $\Lya$ RT equations with constant velocity gradients ($v_{\rm E}/r_0$; with $v_{\rm E}$ being the velocity at the edge $r_0$) for the three geometries. 
By comparing series solutions to Monte Carlo RT simulations, we demonstrate that they work well for small velocity gradients ($v_{\rm E} \lesssim b$) but become inaccurate for large ones ($v_{\rm E} \gtrsim b$), where $b$ is the thermal velocity. 
To characterise spectral features for large velocity gradients, we provide fitting formulae by empirically extending the analytical solution. 
The resulting fitting formulae perform reasonably well up to $v_{\rm E}/b \sim 100$ at large optical depth. 
\end{itemize}

For the ease of use, we provide a summary of the relevant formulae and procedures in Table~\ref{tab:summary_formulae} of Appendix~\ref{sec:summary_formulae}. 

There is much room for further developing the analytical formulae in this study. 
The effects of other physical configurations should be incorporated into $\Lya$ RT equations, including those of clumpy, multiphase gas structures, variations of density, velocity, and temperature, absorptions from dust and molecular hydrogen, and $\Lya$ destructing processes through collisions \citep[e.g.][]{Nebrin_2025}. 
Moreover, the analytical formulae should be closely compared or calibrated to $\Lya$ RT simulations for better understanding of the approximations and their applicable regime. 

The analytical formulae developed in this paper provide important insights for modelling and interpreting $\Lya$ emission under different astrophysical environments. 
For example, $\Lya$ photons can be produced from the star-forming regions and propagate through the interstellar medium in the disk of a galaxy, which broadly corresponds to the slab geometry \citep[e.g.][]{Neufeld_1990}. 
$\Lya$ emission is also expected from gas accretion from the large-scale structure to galaxies through filamentary structures \citep[e.g.][]{Keres_2005}, which resembles the cylindrical geometry. 
Finally, the scattering between $\Lya$ photons and circumgalatic medium, to the zero-th order, follows spherical geometry \citep[e.g.][]{Zheng_2010}. 
In the intergalactic medium, the tidal fields shape the large-scale structure to be sheets, filaments, and knots/voids \citep[e.g.][]{Bond_1996}, whose $\Lya$ emission can be generally represented by the solutions under slab, cylindrical, and spherical geometries, respectively. 
Despite of the similarity in geometry, one should be cautious to directly apply these solutions to real situations given that other gas configurations (e.g. multiphase gas and turbulence) could significantly modify $\Lya$ spectra as well \citep[e.g.][]{Gronke_2016, Nebrin_2025}. 
In the early universe, the spin temperature of 21 centimetre is coupled to the $\HI$ gas temperature through the recoil effect of $\Lya$ RT \citep[e.g.][]{Wouthuysen_1952, Field_1959}, where the formulae under recoil is relevant for interpreting the coupling strength. 
In the end, for cases like strong galactic outflow driven by stellar feedback \citep[e.g.][]{Hopkins_2012} and the Hubble flow from cosmic expansion, the formulae developed under velocity gradients can be helpful. 
Other applications of the formulae include testing numerical codes of $\Lya$ RT \citep[e.g.][]{Zheng_2002a} or implementing sub-grid models in galaxy formation simulations to address $\Lya$ feedback \citep[e.g.][]{Smith_2017, Tomaselli_2021, Kimm_2018, Nebrin_2025, Manzoni_2025}. 
\section*{Acknowledgements}
We thank the referee Olof Nebrin for constructive comments. 
This work is supported by National Science Foundation (NSF) grant AST-2007499. The support and resources from the Center for High Performance Computing at the University of Utah are gratefully acknowledged. 

%%%%%%%%%%%%%%%%%%%%%%%%%%%%%%%%%%%%%%%%%%%%%%%%%%
\section*{Data Availability}
The data underlying this article will be shared on reasonable request to the corresponding author. 
A Python3 package for evaluating the analytical and fitting formulae is provided at \url{https://github.com/PengfeiLiAstro/LyaRTAnalytical.git}. A code for an improved random generator used in Lyman-alpha RT is available at \url{https://github.com/zhengzheng-astro/RandomGenerator}.



\bibliographystyle{mnras}
\bibliography{ref} % if your bibtex file is called example.bib

%\begin{thebibliography}{99}
%\end{thebibliography}

%%%%%%%%%%%%%%%%%%%%%%%%%%%%%%%%%%%%%%%%%%%%%%%%%%

%%%%%%%%%%%%%%%%% APPENDICES %%%%%%%%%%%%%%%%%%%%%

\appendix

\section{An Improved Random Number Generator Used in $\Lya$ RT}\label{sec:randomGenRT}

One step in Monte Carlo simulations of $\Lya$ RT is to generate the
velocity of the hydrogen atom responsible for the scattering. In 
particular, the velocity along the direction of the incident photon follows
the distribution given by the convolution kernel of a Gaussian distribution and
the Lorentz function, 
\begin{equation}
f(u)\propto \frac{e^{-u^2}}{(x-u)^2+a^2},
\end{equation}
where $u$ is in units of the thermal velocity $b=\sqrt{2k_{\rm B}T/m_{\rm H}}$, $a=4.7\times 10^{-4}(T/10^4\, {\rm K})^{-1/2}$ is half of the $\Lya$ natural line 
width in frequency and $x$ is the $\Lya$ line frequency shift, both in units of 
the Doppler frequency width $\Delta\nu_{\rm D}=\nu_\alpha b/c$ (with 
$\nu_\alpha$ the line centre frequency of $\Lya$). 
Note that $a$ is the same as $a_{\rm v}$ used in the main text, and we adopt $a$ in this appendix for simplification.
Given the distribution's symmetry under the transformation of 
$(x,u)\rightarrow (-x,-u)$, we only need to consider the $x\geq 0$ case. 

\begin{figure}
\centering
\includegraphics[alt={A graph showing the targeting distribution function and two of its comparison functions. One is proposed in the previous work, while the other is proposed in this work that tracks the targeting distribution better.}, width=\columnwidth]{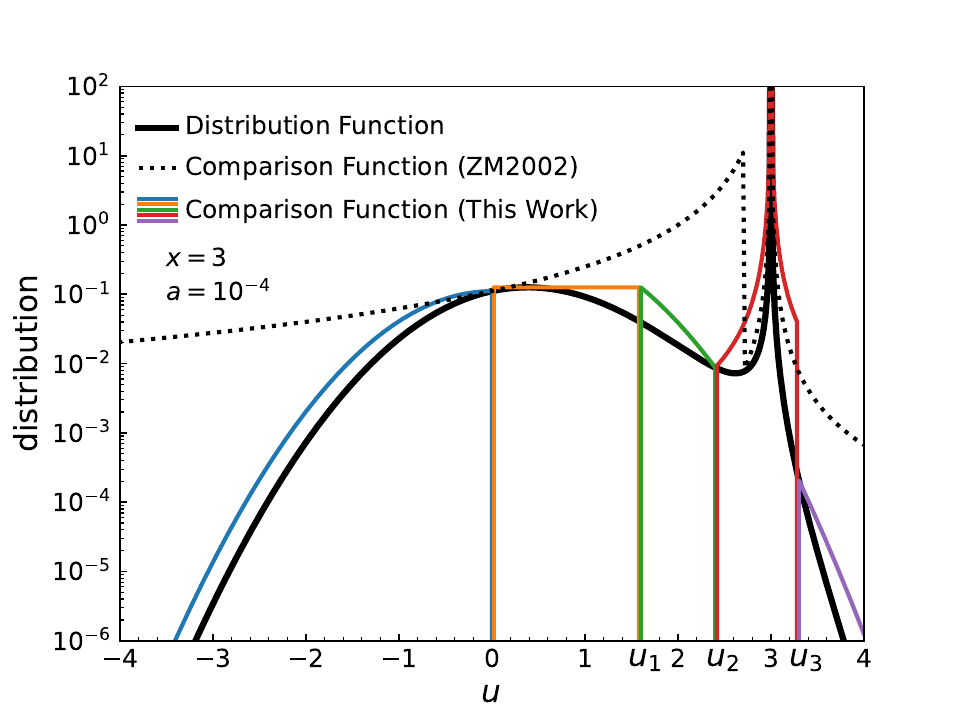}
\caption{ 
Illustration of the distribution function and comparison functions. 
The solid black curve is the distribution function $f(u)$ with $x=3$
and $a=10^{-4}$. The dotted curve is the comparison function proposed
in \citet{Zheng_2002a} and the colour curves are the comparison 
function proposed in this paper. See more details in the text.
}
\label{fig:comparison_func}
\end{figure}

In \citeauthor{Zheng_2002a} (\citeyear{Zheng_2002a}; hereafter ZM2002), an 
algorithm based on the rejection method is developed to draw random numbers 
following the above distribution with given $x$ and $a$. The proposed 
comparison function is
\begin{equation}
g(u)\propto \left\{
\begin{array}{ll}
    1/[(x-u)^2+a^2],           & u\leq u_0 \\
    e^{-u_0^2}/[(x-u)^2+a^2],  & u>u_0.
\end{array}
\right.
\label{eqn:old_compfunc}
\end{equation}
The parameter $u_0$ is chosen to achieve an overall high efficiency of the 
rejection method. While not explicitly mentioned in ZM2002, they adopt 
$u_0=x/(1.01+x/210+x^2/105)$. Note that the distribution $f(u)$ can 
potentially have two peaks, one at $u\sim 0$ and the other at $u\sim x$. 
For small $x$ (e.g. $x\lesssim 1$), only one peak shows up and $f(u)$ is 
sharply peaked around $x$. The comparison function works well for such
a case. For large 
$x$ (e.g. $x\gtrsim 2$), $f(u)$ shows two peaks (see the solid black curve
in Fig.\ref{fig:comparison_func} for an illustration). There is room to 
improve the ZM2002 comparison function (dotted line in 
Fig.\ref{fig:comparison_func}) to enhance the efficiency of the random number
generator.

For $x>2$, we propose the following piecewise function as the comparison 
function,
\begin{equation}
g(u)\propto \left\{
\begin{array}{ll}
    e^{-u^2}/(x^2+a^2),             & u\in (-\infty, 0] \\
    e^{-u_p^2}/[(u_p-x)^2+a^2],     & u\in (0, u_1]\\
    e^{-(u-1/x)^2}e^{1/x^2}/(Ax^2), & u\in (u_1,u_2]\\
    e^{-u_2^2}/[(u-x)^2+a^2],       & u\in (u_2,u_3]\\
    e^{-u^2}/[(u_3-x)^2+a^2],       & u\in (u_3,+\infty).
\end{array}
\right.
\label{eqn:new_compfunc}
\end{equation}
The random number generator for the distribution function in each range can be 
readily implemented, which forms the basis for the rejection method.
The five parameters, $u_p$, $u_1$, $u_2$, $u_3$, and $A$, are determined as 
follows. The parameter $u_p$ is the position of the peak of $f(u)$ near $u=0$. 
By taking the derivative of $f(u)$ with respect to $u$ and setting it to zero, 
we can find a good initial value of $u_p$ to be $x/(x^2+a^2+1)$. A few 
iterations with Newton's root-finding algorithm is enough to obtain the 
converged $u_p$. The values $u_2$ and $u_3$ are supposed to bracket the peak 
position of $f(u)$ near $u=x$, and we take $u_2=px$ and $u_3=qx$ with 
$p=0.7+0.4/x$ and $q=1.1$. By requiring $g(u_2)=f(u_2)$, we solve $A$. Finally,
by requiring $g(u)$ to be continuous at $u=u_1$, we obtain $u_1$.

The coloured curves in Figure~\ref{fig:comparison_func} illustrate the new 
piecewise comparison function in equation~(\ref{eqn:new_compfunc}). 
In comparison with the one in equation~(\ref{eqn:old_compfunc}) used in ZM2002
(dotted curve), 
the new one more closely tracks the shape of $f(u)$ in most 
regions, potentially improving the efficiency of the rejection method of 
generating the $f(u)$ distribution.

The five functions in equation~(\ref{eqn:new_compfunc}) correspond to five 
regions delimited by 0, $u_1$, $u_2$, 
and $u_3$. The method works by first generating a random number $R_1$ uniformly 
distributed in $(0,1)$ to determine which region to use. Then a random number 
$u$ is generated according to the corresponding $g(u)$ in that region. With 
another random number $R_2$ uniformly distributed in $(0,1)$, the random 
number $u$ is kept if $f(u)/g(u)<R_2$ and rejected otherwise. For completeness,
we list the key ingredients below.

The integral of $g(u)$ in each region is calculated as
%\begin{eqnarray}
\begin{equation}
\begin{split}
I_1 =& \int_{-\infty}^0 g(u) du = \frac{\sqrt{\pi}}{2} \frac{1}{x^2+a^2},\\
I_2 =& \int_0^{u_1} g(u) du     = \frac{u_1 e^{-u_p^2}}{(u_p-x)^2+a^2},\\
I_3 =& \int_{u_1}^{u_2} g(u) du = \frac{e^{1/x^2}}{Ax^2} \frac{\sqrt{\pi}}{2} \left[{\rm erf}(u_2-1/x)-{\rm erf}(u_1-1/x)\right],\\
I_4 =& \int_{u_2}^{u_3} g(u) du = e^{-u_2^2}\frac{1}{a} \left(\tan^{-1}\frac{u_3-x}{a}-\tan^{-1}\frac{u_2-x}{a}\right),\\
I_5 =&\int_{u_3}^{+\infty} g(u) du = \frac{1}{(u_3-x)^2+a^2}\frac{\sqrt{\pi}}{2} \left[1-{\rm erf}(u_3)\right].
% \end{eqnarray}
\end{split}
\end{equation}
With $I_{\rm tot}=I_1+I_2+I_3+I_4+I_5$, we define the cumulative fractions, $C_1=I_1/I_{\rm tot}$, $C_2=(I_1+I_2)/I_{\rm tot}$, $C_3=(I_1+I_2+I_3)/I_{\rm tot}$, and $C_4=(I_1+I_2+I_3+I_4)/I_{\rm tot}$. With $R_1$ drawn from the uniform distribution in $(0,1)$, we follow the processes below to draw the random number
$u$:
\begin{itemize}
\item[(1)] if $R_1\leq C_1$, $-u$ is drawn from {\tt PartialExpSq}$(0,+\infty)$;

\item[(2)] if $C_1<R_1\leq C_2$, $u=R u_1$;

\item[(3)] if $C_2<R_1\leq C_3$, $u-1/x$ is drawn from {\tt PartialExpSq}$(u_1-1/x, u_2-1/x)$; 

\item[(4)] if $C_3<R_1\leq C_4$, $(u-x)/a=\tan\left[R\tan^{-1}((u_3-x)/a)\right.+$ $\left.(1-R)\tan^{-1}((u_2-x)/a)\right]$;

\item[(5)] if $R_1> C_4$, $u$ is drawn from {\tt PartialExpSq}$(u_3, +\infty)$,
\end{itemize}
where $R$ is drawn from the uniform distribution in $(0,1)$. The subroutine
{\tt PartialExpSq}$(a_1, a_2)$ returns a random number following the 
distribution $\propto e^{-u^2}$ but only in the range $(a_1, a_2)$\footnote{
For $a_2=+\infty$ and $a_1>0$, a similar algorithm based on the Box–Muller 
transform \citep{Box1958} as implemented in \citet{NR} can be used for 
{\tt PartialExpSq}$(a_1, a_2)$. With a finite $a_2$ and $a_2>a_1>0$, for a
more efficient algorithm, following the Box–Muller-like transform, one can 
first choose a radius and then determine the proper range of the angle.
}.
With the random number $u$ obtained, a random number $R_2$ is drawn from 
the uniform distribution in (0,1). We keep the number $u$ if $f(u)/g(u) < R_2$ and reject it otherwise. 

\begin{figure*}
\centering
\includegraphics[alt={A graph comparing different methods for sampling the targeting distribution function. One panel compares the time for drawing one random number from the targeting distribution. The other panel compares parameter u0 for separating different regimes of one comparison function.}, width=0.45\linewidth]{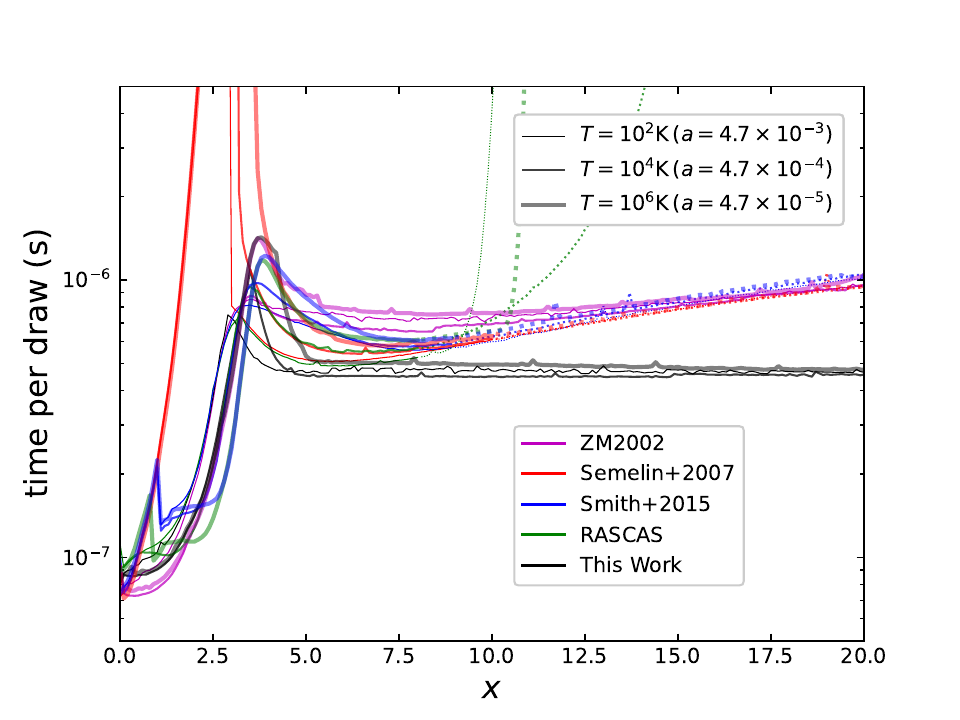}
\includegraphics[width=0.45\linewidth]{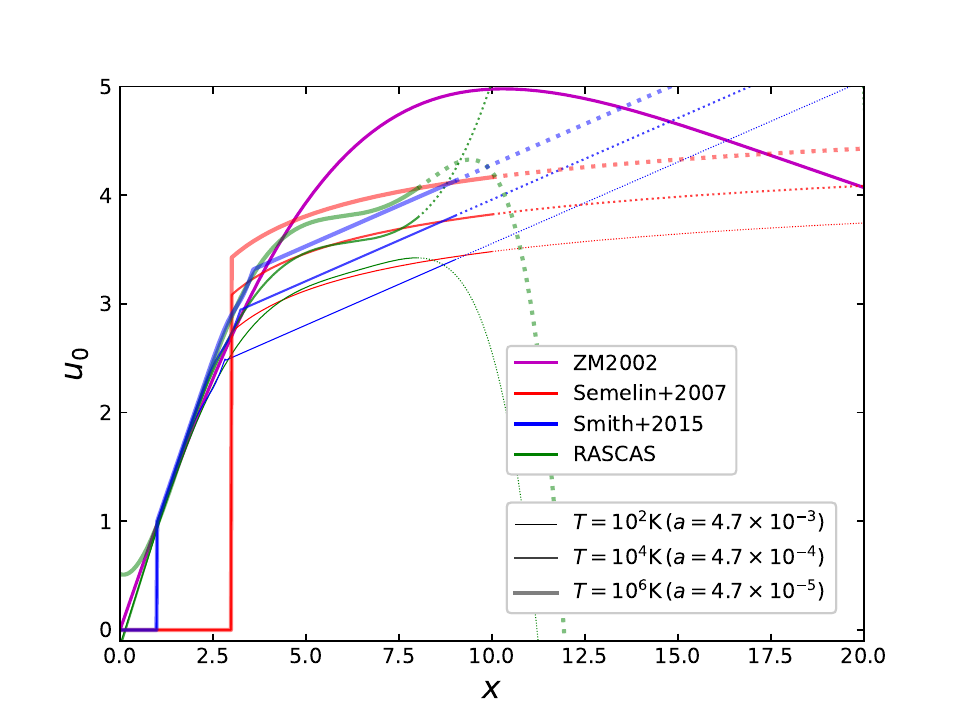}
\caption{
{\it Left}: Performance of generating random numbers following the $f(u)$
distribution. The methods proposed in \citet{Zheng_2002a}, \citet{Semelin2007},
\citet{Smith2015}, \citeauthor{Michel-Dansac2020} (\citeyear{Michel-Dansac2020}; RASCAS code), and this paper are represented by curves of different colours. 
The cases with different temperatures are indicated by the thickness of the 
curves.
{\it Right}: Parameter $u_0$ used in the methods based on the comparison 
function in \citet{Zheng_2002a}. In both panels, the dotted part of a curve 
indicates extrapolation of the $u_0$ function to the range beyond that 
defined in the corresponding paper.
}
\label{fig:runtime_cmp}
\end{figure*}

The comparison function in equation~(\ref{eqn:new_compfunc}) is more complex 
than that in equation~(\ref{eqn:old_compfunc}), the original one in ZM2002.
In terms of performance, there 
are competing factors of improving the comparison function and adding overhead
in the computation. Based on computational tests, we find that a good 
compromise is achieved by using the original ZM2002 method for 
$x<-0.63 \log_{10} a+1.4$ and adopting the new method for higher $x$. We will 
assess the performance of this combination method. While we 
have tried to improve the comparison function in the regime of $x<2$,
we find that the additional overhead makes it hard for improvement. 

In the left panel of Figure~\ref{fig:runtime_cmp}, 
we compare the performance of the
ZM2002 method and the combination method, showing the time per successful 
draw as a function of $x$. We consider three cases for each method, with 
$T=10^2$ K, $10^4$ K, and $10^6$ K, respectively, represented by thin to thick 
curves. The performance of the ZM2002 method (shown with purple curves) is 
already good, $\sim 10^{-7}$s per draw for $x$ close to zero and $\sim 
10^{-6}$ per draw at large $x$ (e.g.  $x\gtrsim 3$ for the $T=10^6$ K case). 
The results are not sensitive to $T$. Our combination method (shown with 
gray curves) improves the performance by a factor of $\sim 2$ in the large 
$x$ regime. 

In the left panel of Figure~\ref{fig:runtime_cmp}, we also show the performance 
of various methods in literature. They all implement the ZM2002 method but with 
different forms of $u_0$. In the right panel of Figure~\ref{fig:runtime_cmp}, 
the corresponding $u_0$ curves are shown. \citet{Semelin2007} use an empirical 
fit\footnote{In the fitting formula of their equation (17), 
\citet{Semelin2007} use both ``log'' and 
``ln'' symbols for logarithm. We interpret ``log'' as natural 
logarithm, too, as we find that this leads to the largest improvement in
performance.} to $u_0$ for $x>3$, 
with dependence on $a$ and $x$, and set 
it to $0$ for $x<3$ (see the red curves in the right panel). 
Since $u_0=0$ is not optimal in the small $x$ regime, their
method (shown with red curves in the left panel) is slower as $x$ shifts away from zero. 
For $T=10^6$ K, the difference from the ZM2002 method in drawing time is more 
than 3 orders of magnitude as $x$ approaches 3. At $x>3$, the performance 
improves substantially, with larger improvement for lower temperature, 
becoming comparable to and better than that of ZM2002. Note that 
at $x>10$ they revert to $u_0=0$ with the distribution truncated to the 
range $[-3, 3]$. In Figure~\ref{fig:runtime_cmp}, we do not show this case 
and instead use dotted curves to represent the situation that their empirical 
fit of $u_0$ continues at $x>10$.

\citet{Smith2015} choose the value of $u_0$ by comparing $x$ with the 
core-to-wing crossover frequency $x_{\rm cw}$, which is about 2.8 for $T=10^2\, \rm K$ 
and 3.9 for $T=10^6$ K. The parameter $u_0$ is effectively 0 for $x<1$ and 
takes different
forms for $1<x<x_{\rm cw}$ and $x_{\rm cw}<x<9$. For $x\gtrsim 9$, $u$ is drawn
from a Gaussian with mean $1/x$. The performance of the \citet{Smith2015} 
method (shown with blue curves in the left panel) is comparable to that of 
ZM2002 and in certain ranges of $x$ it is better. Note that we simply 
extrapolate their $u_0$ form for $x_{\rm cw}<x<9$ to $x\gtrsim 9$ (shown with
blue dotted curves in both panels).

\citet{Michel-Dansac2020} use an empirical fitting function of $u_0(x,a)$ in their RASCAS code for $x<8$. At higher values of $x$, they follow \citet{Smith2015} to draw $u$ from a Gaussian with mean $1/x$. Their method (green curves in the
left panel) has a slightly better performance at $x\gtrsim 3$ than that of \citet{Smith2015}\footnote{In Fig.4 of \citet{Michel-Dansac2020}, the RASCAS method appears to 
perform much better (e.g. by more than one order of magnitude for 
$a\leq 10^{-3}$) than that of \citet{Smith2015}. 
We find that this may result from using base-10 rather than 
natural logarithm to compute the core-to-wing crossover frequency $x_{\rm cw}$ 
in eq.(21) of \citet{Smith2015}.
}.

The $u_0$ curves in the right panel of Figure~\ref{fig:runtime_cmp} can help
understand the performance comparison based on the ZM2002 method. 
The curves adopted in ZM2002, \citet{Smith2015}, and \citet{Michel-Dansac2020} 
are almost on top of each other at $x\lesssim 2.5$, resulting in similar 
performance. At $3\lesssim x \lesssim 9$, the $u_0$ curves in 
\citet{Semelin2007}, \citet{Smith2015}, and \citet{Michel-Dansac2020} are more
or less close to each other, and we see similar performance of the three 
method, slightly better than that of ZM2002. 

At $x\gtrsim 4$, the method we propose here has the best performance, 
although not by a large factor. Given the large fraction of overhead with 
this method, the performance is impressive. There can be room to fine tune 
the subroutines related to the new comparison function to further reduce 
the overhead. Given the reasonably effective performance seen here, we make no 
further optimization. 

The code of the improved random generator for the $f(u)$ distribution will be provided at \url{https://github.com/zhengzheng-astro/RandomGenerator}.

\section{Alternative form of $\Lya$ RT solutions under cylindrical geometry}\label{sec:spec_additional_derivation}
In this appendix, we provide an alternative form of $\Lya$ RT solutions for a static, uniform cloud under the cylindrical geometry. 
Unlike the slab and spherical geometries, $\Lya$ spectra under cylindrical geometry do not have a closed-form solution in general. 
However, the functional form can still be well informed with the help of Lerch transcendent as shown below. 

We start from the series solution in equation~(\ref{eq:ses_full}), setting $\tau_{\rm s}=0$ and $y_{\rm i}=0$. 
The term $J_0(\lambda_n \tau_{\rm s})$ becomes 1, while $J_1(\lambda_n\tau_0)$ can be approximated using its asymptotic form in equation~(\ref{eq:J1_approx}). 
This approximation leads to less than 2 per cent of error on $J_1(\lambda_n\tau_0)$ for all orders of $\lambda_n$. 
With these steps, the summation part of equation~(\ref{eq:ses_full}) can be rewritten as 
\begin{equation}
    S = \frac{\pi}{\sqrt{2}}\sum_{n=0}^{\infty}(-1)^n \left(n + \frac{3}{4} \right)^{1/2}\exp \left[-\left( n+\frac{3}{4}\right) \tilde{y} \right], 
    \label{eq:apdx_sn}
\end{equation}
where $\tilde{y} \equiv \pi |y|/\tau_0$. 
This summand in equation~(\ref{eq:apdx_sn}) can be substituted by the derivative of its primitive function, after which we interchange the summation and the differentiation to obtain 
\begin{equation}
\begin{split}
    S = &  -\frac{\pi}{\sqrt{2}}\frac{\rm d}{{\rm d}\tilde{y}} \sum_{n=0}^{\infty} (-1)^n \frac{\exp \left[-\left( n+\frac{3}{4}\right) \tilde{y} \right]}{\left(n + \frac{3}{4} \right)^{1/2}} \\
    = & -\frac{\pi}{\sqrt{2}}\frac{\rm d}{{\rm d}\tilde{y}} \sum_{k=0}^{\infty} \left\{ \frac{\exp \left[-\left( 2k+\frac{3}{4}\right) \tilde{y} \right]}{\left(2k + \frac{3}{4} \right)^{1/2}} - \frac{\exp \left[-\left( 2k+\frac{7}{4}\right) \tilde{y} \right]}{\left(2k + \frac{7}{4} \right)^{1/2}}\right\}. 
\end{split}
    \label{eq:apdx_sk}
\end{equation}
The second step further splits the summation into two for even and odd orders of $n$. 
The summand can be rewritten adopting the Lerch transcendent 
\begin{equation}
    \Phi(z, s, \alpha) = \sum_{k=0}^{\infty} \frac{z^k}{(k+\alpha)^s} =  \frac{1}{\Gamma(s)}\int_0^\infty {\rm d}t\frac{t^{s-1} e^{-\alpha t}}{1-ze^{-t}}, 
    \label{eq:apdx_Lerch}
\end{equation}
where $\Gamma(s)$ is the gamma function with $s=1/2$. 
Replacing the summation terms in equation~(\ref{eq:apdx_sk}) with the integral form of Lerch transcendent, after some algebra, we arrive at 
\begin{equation}
    \begin{split}
    S = & -\sqrt{\pi} \frac{{\rm d}}{{\rm d \tilde{y}}} 
    \int_0^\infty {\rm d}t \frac{\exp\left( -\frac{3}{4}\tilde{y}-\frac{3}{8} t^2\right)}{1+\exp\left( -\tilde{y}-\frac{1}{2} t^2\right)} \\
    = & \sqrt{\pi} \exp\left( -\frac{3}{4}\tilde{y} \right) \int_0^\infty {\rm d}t 
    \frac{\frac{3}{4}\exp\left(-\frac{3}{8} t^2\right)
    -\frac{1}{4} \exp\left( -\tilde{y}-\frac{7}{8} t^2\right)}{ \left[ 1+\exp\left( -\tilde{y}-\frac{1}{2} t^2\right) \right]^2}, 
    \end{split}
    \label{eq:apdx_sk_simplified}
\end{equation}
where the second step interchanges the differentiation and integration, and carries out the derivative with respect to $\tilde{y}$. 
The integral in the second step can not be analytically evaluated, unfortunately. 
But its limiting behaviour is still interesting. 
At $\tilde{y} = 0$, the function $S(\tilde{y})$ in equation~(\ref{eq:apdx_sk_simplified}) converges to $\sim 0.661$.\footnote{The summation in equation~(\ref{eq:apdx_sn}) diverges at $\tilde{y}=0$, but the limit exists as $\tilde{y}$ approaches zero. 
See similar discussions in \cite{Smith_2025}. }
As $\tilde{y} \to \infty$, the function $S(\tilde{y})$ follows $\left( \sqrt{6}\pi/4 \right) \exp \left( - 3\tilde{y}/4 \right) - \left( \sqrt{14}\pi/7 \right) \exp \left( -7 \tilde{y}/4 \right)$ by setting $\exp(-\tilde{y})=0$ in the denominator of the integrand in equation~(\ref{eq:apdx_sk_simplified}), consistent with the expectation from equation~(\ref{eq:apdx_sn}) by only keeping the first two terms in the summation. 

The limiting behaviour enlightens us to come up with closed-form functions mimicking the true $S(\tilde{y})$ in equation~(\ref{eq:apdx_sk_simplified}). 
The following formulae are adopted to achieve the goal, 
\begin{equation}
    \begin{split}
    & \int_0^\infty {\rm d}t \frac{\exp\left( -\frac{3}{8} t^2\right)}{ \left[ 1+\exp\left( -\tilde{y}-\frac{1}{2} t^2\right) \right]^2} \simeq  \frac{p_\infty + (4p_0 - p_\infty)\exp(-\tilde{y})}{[1 +\exp(-\tilde{y})]^2}, \\
    & \int_0^\infty {\rm d}t \frac{\exp\left( -\frac{7}{8} t^2\right)}{ \left[ 1+\exp\left( -\tilde{y}-\frac{1}{2} t^2\right) \right]^2} \simeq  \frac{q_\infty + (4q_0 - q_\infty)\exp(-\tilde{y})}{[1 +\exp(-\tilde{y})]^2} ,
    \label{eq:apdx_int}
    \end{split}
\end{equation} 
where the parameters $p_0$ and $q_0$, $p_\infty$ and $q_\infty$ are set by the integral limits when $\tilde{y} \to 0$ and $\infty$, respectively, with $p_0 \simeq 0.601$, $q_0 \simeq 0.310$, $p_\infty=\sqrt{6\pi}/3 \simeq 1.45$,  and $q_\infty = \sqrt{14\pi}/7 \simeq 0.947$. 
The error of such approximation is less than 2 per cent for both cases in equation~(\ref{eq:apdx_int}). 
Combining equation~(\ref{eq:ses_full}), (\ref{eq:apdx_sn}), (\ref{eq:apdx_sk_simplified}), and (\ref{eq:apdx_int}), the $\Lya$ spectrum has the following form 
\begin{equation}
    \begin{split}
    J_{\rm cyl} (y) = & \frac{1}{2\pi r_0} \frac{\sqrt{6}}{12 \pi \tau_0 \voigt} \frac{1}{ 1+ \cosh\left( -\frac{\pi |y|}{\tau_0}\right)} \\
    \times & \left[ 1.92\exp \left(\frac{\pi|y|}{4\tau_0} \right) + 0.85\exp \left(-\frac{3\pi|y|}{4\tau_0} \right) \right. \\
    & \left. - 0.13\exp \left(-\frac{7\pi|y|}{4\tau_0} \right) \right], 
    \end{split}
    \label{eq:apdx_ana_Lerch}
\end{equation}
which has identical functional forms as equation~(\ref{eq:cyl_full}) at $\tau_{\rm s}=0$ and $x_{\rm i}=0$, while the coefficients in front of the exponential terms are slightly different. 
This difference is mainly caused by different fitting procedures, where the coefficients in equation~(\ref{eq:cyl_full}) are constrained by function values at multiple $|y|$, while the ones in equation~(\ref{eq:apdx_ana_Lerch}) are constrained by function values at $|y| \to 0$ and $\infty$. 

To further obtain $\Lya$ spectra when $\tau_{\rm s} \neq 0$, one can expand the function $J_0(\lambda_n \tau_{\rm s})$ in equation~(\ref{eq:ses_full}) following the polynomial representation of Bessel function  $J_0(z) = \sum_{m=0}^{\infty} (-1)^m (z/2)^{2m}/(m!)^2$. 
We present an example for the second term ($m=1$) of the expansion, $-(\lambda_n \tau_{\rm s})^2/4$. 
By replacing $J_0(\lambda_n \tau_{\rm s})$ with $-(\lambda_n \tau_{\rm s})^2/4$ and $J_1(\lambda_n \tau_0)$ with its asymptotic form (equation~(\ref{eq:J1_approx})), the summation part in equation~(\ref{eq:ses_full}) becomes 
\begin{equation}
    S_{m=1} = -\frac{\pi}{4\sqrt{2}} \left(\frac{\tau_{\rm s}}{\tau_0} \right)^2\sum_{n=0}^{\infty}(-1)^n \left(n + \frac{3}{4} \right)^{5/2}\exp \left[-\left( n+\frac{3}{4}\right) \tilde{y} \right]. 
    \label{eq:apdx_S1}
\end{equation}
Applying similar steps shown in equation~(\ref{eq:apdx_sk}) yields  
\begin{equation}
    S_{m=1} = \frac{\pi}{4\sqrt{2}} \left(\frac{\tau_{\rm s}}{\tau_0} \right)^2 \frac{{\rm d}^3}{{\rm d} \tilde{y}^3} \sum_{n=0}^{\infty}(-1)^n \frac{ \exp \left[-\left( n+\frac{3}{4}\right) \tilde{y} \right]}{\left(n + \frac{3}{4} \right)^{1/2}}. 
    \label{eq:apdx_S1_dev}
\end{equation}
Afterwards, the Lerch transcendent can be applied in similar ways to the case of $\tau_{\rm s}=0$ (equation~(\ref{eq:apdx_sk_simplified})), followed by the approximations resembling equation~(\ref{eq:apdx_int}) if a closed-form function is desired. 
With this procedure, additional higher-order terms of $\tau_{\rm s}/\tau_0$ can be included in analogous forms to equation~(\ref{eq:apdx_S1_dev}), but with higher-order derivatives with respect to $\tilde{y}$. 
Given that an applicable fitting function is already presented in equation~(\ref{eq:cyl_full}), we do not carry out the detailed calculations for this method and leave them to interested readers. 
\section{Fitting formula for $\Lya$ spectra under velocity gradient}\label{sec:vgrad_fitfunc}

\begin{table*}
	\centering
	\caption{Dependence of parameters $J_{\rm A}$, $c_\lambda$, $n_\lambda$, $x_\gamma$, $n_\gamma$, and $c_\gamma$ in equation~(\ref{eq:ana_central_ext_pla}), (\ref{eq:ana_central_ext}), and (\ref{eq:ana_central_ext_sph}) on physical parameters $\log(v_{\rm E}/b)$ and $\tau_0$. 
    The three columns correspond to slab, cylindrical, and spherical geometries, respectively. 
    }
	\label{tab:fitfunc_geo}
	\begin{tabular}{l l l l} % four columns, alignment for each
		\hline \hline
		& Slab & Cylindrical & Spherical \\
		\hline 
        \begin{minipage}[c][13em][c]{0.2cm}
        $J_{\rm A}$
        \end{minipage}
        & 
        \begin{minipage}[c][13em][c]{5.7cm}
        \centering
        \[
        \begin{split}
            & \frac{0.588}{1+ \left( \frac{|v_{\rm E}/b|}{3.06} \right)^{1.24}} \\
                \times & \left\{1+ \left( \frac{|v_{\rm E}/b|}{3.06} \right)^{0.00208} \right. \\
            & + \left. \left[ \frac{|v_{\rm E}/b|}{130 \left( \frac{a_v \tau_0}{1.49 \times 10^4} \right)^{0.427}} \right]^{4.18} \right\}
        \end{split}
        \]
        \end{minipage}
        & 
        \begin{minipage}[c][13em][c]{5.7cm}
        \centering
        \[
        \begin{split}
            & \frac{0.578}{1+ \left( \frac{|v_{\rm E}/b|}{2.09} \right)^{1.06}} \\
            \times & \left\{1+ \left( \frac{|v_{\rm E}/b|}{2.09} \right)^{0.000214} \right. \\
            & + \left. \left[ \frac{|v_{\rm E}/b|}{140 \left( \frac{a_v \tau_0}{1.49 \times 10^4} \right)^{0.389}} \right]^{3.99} \right\}
        \end{split}
        \]
        \end{minipage}
        & 
        \begin{minipage}[c][13em][c]{5.7cm}
        \centering
        \[
        \begin{split}
            & \frac{0.616}{1+ \left( \frac{|v_{\rm E}/b|}{1.76} \right)^{1.03}} \\
            \times & \left\{1+ \left( \frac{|v_{\rm E}/b|}{1.76} \right)^{0.000141} \right. \\
            & + \left. \left[ \frac{|v_{\rm E}/b|}{123 \left( \frac{a_v \tau_0}{1.49 \times 10^4} \right)^{0.38}} \right]^{4.35} \right\}
        \end{split}
        \]
        \end{minipage}
        \\
        \hline
        \begin{minipage}[c][10em][c]{0.2cm}
        $c_\lambda$ 
        \end{minipage}
        & 
        \begin{minipage}[c][10em][c]{5.7cm}
        \centering
        \[ 
        \begin{split}
            & \left[9.04 + 2.72 \log \left( \frac{a_{\rm v} \tau_0}{1.49 \times 10^4} \right) \right] \\
            \times & \left\{ 1+\left[ \frac{|v_{\rm E}/b|}{15 \left( \frac{a_{\rm v} \tau_0}{1.49 \times 10^4} \right)^{0.448}} \right] 
            ^{0.495\log \left( \frac{a_{\rm v} \tau_0}{1.49 \times 10^4} \right) +2.24} \right\}
        \end{split}
        \]
        \end{minipage}
        & 
        \begin{minipage}[c][10em][c]{5.7cm}
        \centering
        \[ 
        \begin{split}
            & \left[8.25 + 2.31 \log \left( \frac{a_{\rm v} \tau_0}{1.49 \times 10^4} \right) \right] \\
            \times & \left\{ 1 + \left[\frac{|v_{\rm E}/b|}{13.4 \left( \frac{a_{\rm v} \tau_0}{1.49 \times 10^4} \right)^{0.437}} \right]
            ^{0.512\log \left( \frac{a_{\rm v} \tau_0}{1.49 \times 10^4} \right) + 2.4} \right\}
        \end{split}
        \]
        \end{minipage}
        & 
        \begin{minipage}[c][10em][c]{5.7cm}
        \centering
        \[ 
        \begin{split}
            & \left[8.3 + 2.33 \log \left( \frac{a_{\rm v} \tau_0}{1.49 \times 10^4} \right) \right] \\
            \times & \left\{ 1 + \left[\frac{|v_{\rm E}/b|}{11.8 \left( \frac{a_{\rm v} \tau_0}{1.49 \times 10^4} \right)^{0.405}} \right]
            ^{0.495\log \left( \frac{a_{\rm v} \tau_0}{1.49 \times 10^4} \right) + 2.39} \right\}
        \end{split}
        \]
        \end{minipage}
        \\
        \hline 
        \begin{minipage}[c][6em][c]{0.2cm}
        $n_\lambda$ 
        \end{minipage}
        &
        \begin{minipage}[c][6em][c]{5.7cm}
        \centering
        \[ 
        \begin{split}
            \frac{2.4}{1 - \left[ \frac{|v_{\rm E}/b|}{(a_{\rm v} \tau_0)^{0.402}} \right]^{1.42} + 1.82 \left[ \frac{|v_{\rm E}/b|}{(a_{\rm v} \tau_0)^{0.402}} \right]^{1.18}}
            \end{split}
        \]
        \end{minipage}
        &
        \begin{minipage}[c][6em][c]{5.7cm}
        \centering
        \[ 
        \begin{split}
            \frac{2.41}{1 - \left[ \frac{|v_{\rm E}/b|}{(a_{\rm v} \tau_0)^{0.427}} \right]^{1.27} + 2.23 \left[ \frac{|v_{\rm E}/b|}{(a_{\rm v} \tau_0)^{0.427}} \right]^{1.08}}
            \end{split}
        \]
        \end{minipage}
        &
        \begin{minipage}[c][6em][c]{5.7cm}
        \centering
        \[ 
        \begin{split}
            \frac{2.39}{1 - \left[ \frac{|v_{\rm E}/b|}{(a_{\rm v} \tau_0)^{0.415}} \right]^{1.16} + 2.27 \left[ \frac{|v_{\rm E}/b|}{(a_{\rm v} \tau_0)^{0.415}} \right]^{1.03}}
            \end{split}
        \]
        \end{minipage}
        \\
        \hline
        \begin{minipage}[c][8em][c]{0.2cm}
        $x_\gamma$ 
        \end{minipage}
        &
        \begin{minipage}[c][8em][c]{5.7cm}
        \centering
        \[ 
        \begin{split}
             & \frac{1}{1 + 3.89 \left[ \frac{|v_{\rm E}/b|}{ (a_{\rm v} \tau_0)^{0.237}} \right]^{1.39}} \\
             + & 0.0101 \left( \frac{a_{\rm v} \tau_0}{1.49 \times 10^4} \right)^{-0.614}
        \end{split}
        \]
        \end{minipage}
        &
        \begin{minipage}[c][8em][c]{5.7cm}
        \centering
        \[ 
        \begin{split}
             & \frac{0.986}{1 + 1.82 \left[ \frac{|v_{\rm E}/b|}{ (a_{\rm v} \tau_0)^{0.114}} \right]^{1.38}} \\
             + & 0.02 \left( \frac{a_{\rm v} \tau_0}{1.49 \times 10^4} \right)^{-0.235}
        \end{split}
        \]
        \end{minipage}
        &
        \begin{minipage}[c][8em][c]{5.7cm}
        \centering
        \[ 
        \begin{split}
             & \frac{0.895}{1 + 1.35 \left[ \frac{|v_{\rm E}/b|}{ (a_{\rm v} \tau_0)^{0.0303}} \right]^{1.21}} \\
             + & 0.152 \left( \frac{a_{\rm v} \tau_0}{1.49 \times 10^4} \right)^{0.0407}
        \end{split}
        \]
        \end{minipage}
        \\
        \hline
        \begin{minipage}[c][6em][c]{0.2cm}
        $n_\gamma$ 
        \end{minipage}
        &
        \begin{minipage}[c][6em][c]{5.7cm}
        \centering
        \[ 
        \begin{split}
            \frac{8.14 \frac{|v_{\rm E}/b|}{(a_{\rm v} \tau_0)^{0.372}}}{1+3.77 \frac{|v_{\rm E}/b|}{(a_{\rm v} \tau_0)^{0.372}}}
        \end{split}
        \]
        \end{minipage}
        &
        \begin{minipage}[c][6em][c]{5.7cm}
        \centering
        \[ 
        \begin{split}
            \frac{8.28 \frac{|v_{\rm E}/b|}{(a_{\rm v} \tau_0)^{0.312}}}{1+3.95 \frac{|v_{\rm E}/b|}{(a_{\rm v} \tau_0)^{0.312}}}
        \end{split}
        \]
        \end{minipage}
        &
        \begin{minipage}[c][6em][c]{5.7cm}
        \centering
        \[ 
        \begin{split}
            \frac{16.8 \frac{|v_{\rm E}/b|}{(a_{\rm v} \tau_0)^{0.349}}}{1+10.2 \frac{|v_{\rm E}/b|}{(a_{\rm v} \tau_0)^{0.349}}}
        \end{split}
        \]
        \end{minipage}
        \\
        \hline
        \begin{minipage}[c][8em][c]{0.2cm}
        $c_\gamma$ 
        \end{minipage}
        &
        \begin{minipage}[c][8em][c]{5.7cm}
        \centering
        \[ 
        \begin{split}
            & \left[ -1.32 - 0.661\log \left( \frac{a_{\rm v} \tau_0}{1.49 \times 10^4} \right) \right] \\
            \times & \exp \left[-\frac{|\log (|v_{\rm E}/b|)  - 1.67|^{0.781}}{0.319} \right]
        \end{split}
        \]
        \end{minipage}
        &
        \begin{minipage}[c][8em][c]{5.7cm}
        \centering
        \[ 
        \begin{split}
            & \left[ -1.83 - 0.537\log \left( \frac{a_{\rm v} \tau_0}{1.49 \times 10^4} \right) \right] \\
            \times & \exp \left[-\frac{|\log (|v_{\rm E}/b|)  -1.61|^{4.19}}{0.559} \right]
        \end{split}
        \]
        \end{minipage}
        &
        \begin{minipage}[c][8em][c]{5.7cm}
        \centering
        \[ 
        \begin{split}
            & \left[ -2.35 - 0.268\log \left( \frac{a_{\rm v} \tau_0}{1.49 \times 10^4} \right) \right] \\
            \times & \exp \left[-\frac{|\log (|v_{\rm E}/b|)  -1.84|^{4.22}}{1.85} \right]
        \end{split}
        \]
        \end{minipage}
        \\
		\hline
    \end{tabular}
    \raggedright
    {\it Note.} The applicable range of these formulae spans $\log(v_{\rm E}/b)=(-1, 2)$ and $\log (a_{\rm v}\tau_0) = (2.2, 4.2)$ (i.e. $\log \tau_0=(4, 6)$ at $T=10\, \rm K$). 
    The uncertainty of reproducing simulated $\Lya$ spectra with these formulae is presented in Fig.~\ref{fig:err_fitfunc}. 
\end{table*}

\begin{figure*}
    \includegraphics[alt={A graph comparing the uncertainties between the fitting formulae and the series solutions for evaluating Lyman-alpha spectra under non-zero velocity gradients. The comparison is done for various optical depths, velocity gradients, and geometries. Both the fitting formulae and the series solution perform well for small velocity gradients, while the fitting formulae perform better than the series solutions for large velocity gradients under large optical depths.}, width=\textwidth]{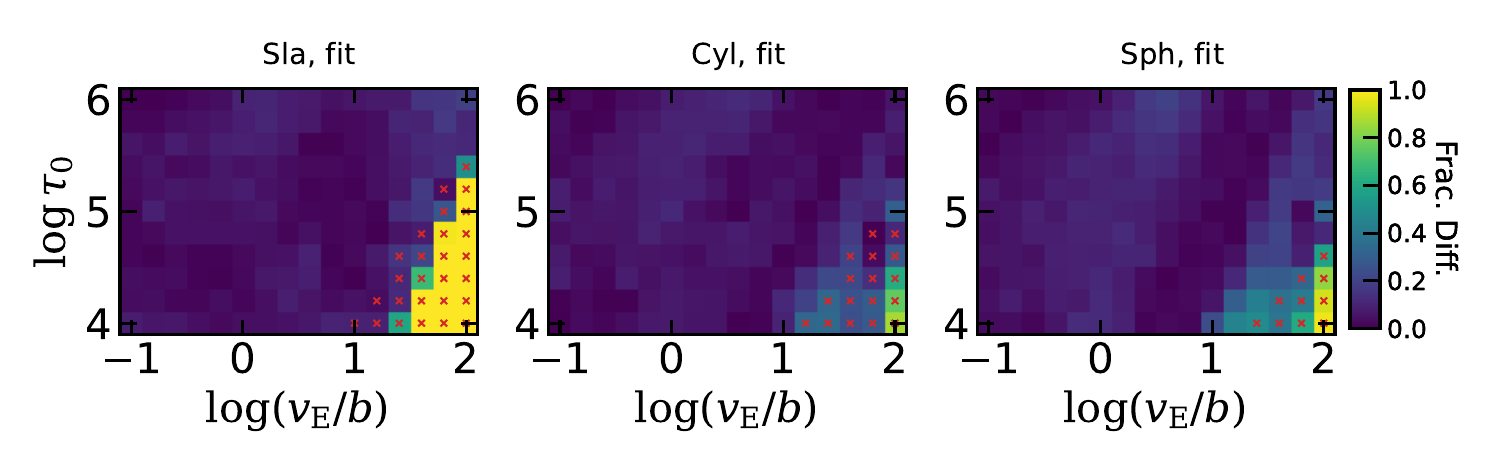}
    \includegraphics[width=\textwidth]{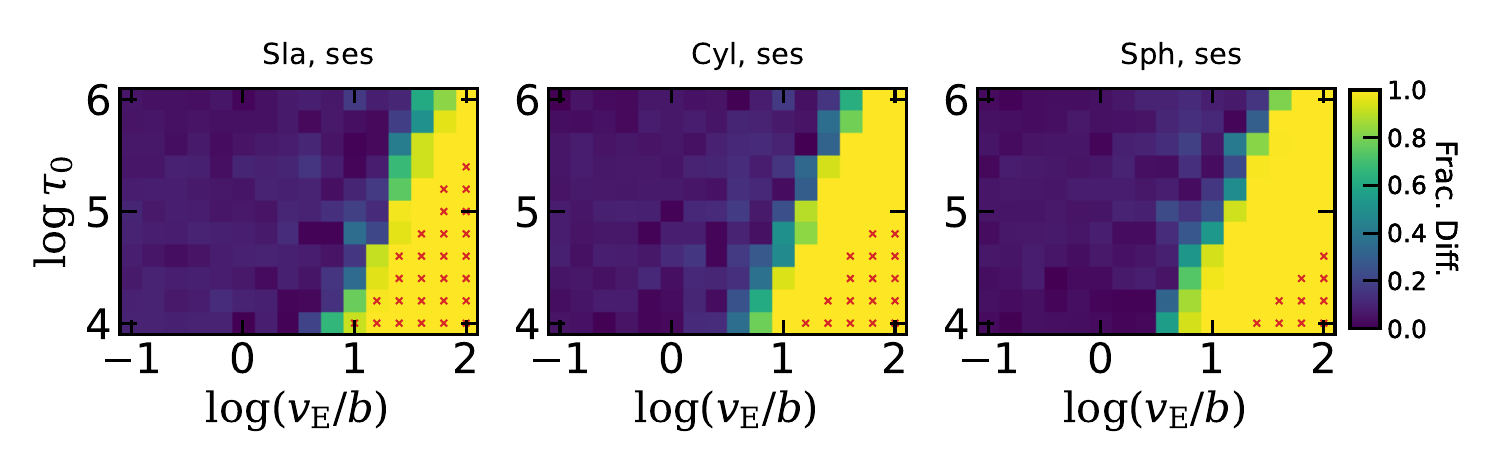}
    \caption{\textit{Top}: uncertainty of the fitting formulae, equation~(\ref{eq:ana_central_ext_pla}), (\ref{eq:ana_central_ext}), and (\ref{eq:ana_central_ext_sph}), for different optical depths and velocity gradients. 
    The uncertainty is estimated as the fractional difference between the $\Lya$ spectrum from the fitting formulae and that from $\Lya$ RT simulations, with respect to that from $\Lya$ RT simulations, at the frequency where the simulated $\Lya$ spectrum has the maximal flux. 
    The three columns correspond to slab, cylindrical, and spherical geometries, respectively. 
    The crosses label the model parameters whose corresponding spectra are not adopted in the the joint fitting as discussed in Appendix~\ref{sec:vgrad_fitfunc}. 
    \textit{Bottom}: same as the top row but for series solutions, equation~(\ref{eq:sesv_full_sla}), (\ref{eq:sesv_full}), and (\ref{eq:sesv_full_sph}). 
    }
    \label{fig:err_fitfunc}
\end{figure*}

In this appendix, we determine the fitting formulae for calculating $J_{\rm A}$, $c_\lambda$, $n_\lambda$, $x_\gamma$, $n_\gamma$, and $c_\gamma$ from physical parameters $\tau_0$ and $v_{\rm E}/b$, for each geometry separately as mentioned in Section~\ref{subsec:fitting_vgrad}. 
By combining these fitting formulae with equation~(\ref{eq:ana_central_ext}), (\ref{eq:ana_central_ext_pla}), and (\ref{eq:ana_central_ext_sph}), we can analytically generate $\Lya$ spectra spanning a wide range of optical depths and velocity gradients for each geometry. 

We use cylindrical geometry to introduce the basic procedures for obtaining the fitting formulae, with the same procedure applied to slab and spherical geometries. 
First, we run RT simulations on a uniform grid of $\log (v_{\rm E}/b)$ from -1 to 2 with $\Delta \log (v_{\rm E}/b)=0.2$ and $\log \tau_0$ from 4 to 6 with $\Delta \log \tau_0 =0.2$. 
Next, we use the formula in equation~(\ref{eq:ana_central_ext}) to fit each simulated $\Lya$ spectrum from this grid, and collect the best-fit parameters of $J_{\rm A}$, $c_\lambda$, $n_\lambda$, $x_\gamma$, $n_\gamma$, and $c_\gamma$. 
Then, we empirically construct the fitting formulae by observing how each best-fit parameter varies with $\tau_0$ and $v_{\rm E}/b$. 
Finally, we determine the parameters in the fitting formulae by jointly fitting all simulated $\Lya$ spectra on the grid. 
Certain spectra on the grid are omitted in the joint fitting because their spectral features can not be faithfully reproduced by the functional form in equation~(\ref{eq:ana_central_ext}). 
Their corresponding parameters are labelled by crosses in Fig.~\ref{fig:err_fitfunc}, preferentially located at the corner of large velocity gradients and small optical depths. 
The fitting formulae for slab and spherical geometries are obtained similarly. 

Table~\ref{tab:fitfunc_geo} presents fitting formulae for $J_{\rm A}$, $c_\lambda$, $n_\lambda$, $x_\gamma$, $n_\gamma$, and $c_\gamma$ for each of the three geometries. 
To evaluate their accuracy of reproducing simulated $\Lya$ spectra, we calculate the fractional difference between a $\Lya$ spectrum from fitting formulae and that from $\Lya$ RT simulations, with respect to that from $\Lya$ RT simulations, at the frequency where the simulated $\Lya$ spectrum has the maximal flux. 
The results are shown in the top row of Fig.~\ref{fig:err_fitfunc} for different $v_{\rm E}/b$, $\tau_0$, and geometries. 
The fractional difference is smaller than 10 per cent for most parts of the parameter space, while becomes larger towards large $v_{\rm E}/b$ and small $\tau_0$. 
As a comparison between series solutions and fitting formulae, we present the same results for the series solutions in the bottom row of Fig.~\ref{fig:err_fitfunc}. 
The fitting formulae show improvement across the parameter space, particularly for $v_{\rm E}/b \sim 100$ under large optical depth. 
\section{Summary of analytical formulae}\label{sec:summary_formulae}

\begin{table*}
	\centering
	\caption{Summary of the analytical formulae for $\Lya$ spectra $J$ from a uniform gas cloud with zero or constant velocity gradients under slab, cylindrical, and spherical geometries. 
    The heading of each row lists the dependence of the analytical formulae on source positions $\tau_{\rm s}$, initial frequency $x_{\rm i}$, and ratio of the edge velocity to the thermal velocity, $v_{\rm E}/b$. 
    The origins of the functional forms are also included, being closed-form solutions, series solutions, or fitting formulae. 
    The analytical procedure for incorporating the recoil effect into $\Lya$ spectra is described in the bottom row of the table. 
    }
	\label{tab:summary_formulae}
	\begin{tabular}{l l l l} % four columns, alignment for each
		\hline \hline
		& Slab & Cylindrical & Spherical \\
		\hline 
        \begin{minipage}[c][18em][c]{1.5cm}
        $\tau_{\rm s} = 0$\\$x_{\rm i} = 0$\\$v_{\rm E}/b=0$\\Closed-form
        \end{minipage}
        & 
        \begin{minipage}[c][18em][c]{5.2cm}
        \centering
        \[
        \begin{split}
            & \frac{1}{2} \left( \frac{\sqrt{6}x^2}{12 \sqrt{\pi} a_{\rm v}\tau_0} \right)
             \left/ \left[ 1 + \cosh \left( \sqrt{ \frac{2\pi^3}{27}} \frac{ \left|x^3 \right|}{a_{\rm v}\tau_0} \right) \right]\right.\\
            & \times 2\cosh \left( \frac{1}{2} \sqrt{ \frac{2\pi^3}{27}} \frac{ \left|x^3 \right|}{a_{\rm v}\tau_0} \right)
        \end{split}
        \]
        \end{minipage}
        & 
        \begin{minipage}[c][18em][c]{5.2cm}
        \centering
        \[
        \begin{split}
            & \frac{1}{2\pi r_0} \left( \frac{\sqrt{6}x^2}{12 \sqrt{\pi} a_{\rm v}\tau_0} \right)
             \left/ \left[ 1 + \cosh \left( \sqrt{ \frac{2\pi^3}{27}} \frac{ \left|x^3 \right|}{a_{\rm v}\tau_0} \right) \right]\right. \\
            & \times \left[ 1.92\exp \left(\frac{1}{4}\sqrt{ \frac{2\pi^3}{27}} \frac{ \left|x^3 \right|}{a_{\rm v}\tau_0}  \right) \right. \\
            & \,\,\,\,\,\, + 1.32\exp \left(-\frac{3}{4} \sqrt{ \frac{2\pi^3}{27}} \frac{ \left|x^3 \right|}{a_{\rm v}\tau_0} \right) \\
            & \,\,\,\,\,\, -\left. 0.609\exp \left(-\frac{7}{4} \sqrt{ \frac{2\pi^3}{27}} \frac{ \left|x^3 \right|}{a_{\rm v}\tau_0} \right) \right]
        \end{split}
        \]
        \end{minipage}
        & 
        \begin{minipage}[c][18em][c]{5.2cm}
        \centering
        \[
        \begin{split}
            & \frac{1}{4\pi r_0^2}\left( \frac{\sqrt{6}x^2}{12 \sqrt{\pi} a_{\rm v}\tau_0} \right)
             \left/ \left[ 1 + \cosh \left( \sqrt{ \frac{2\pi^3}{27}} \frac{ \left|x^3 \right|}{a_{\rm v}\tau_0} \right) \right]\right. \\
            & \times \pi
        \end{split}
        \]
        \end{minipage}
        \\
        \hline
        \begin{minipage}[c][5em][c]{1.5cm}
        $\tau_{\rm s} \neq 0$\\ $x_{\rm i} \neq 0$\\$v_{\rm E}/b=0$\\Closed-form
        \end{minipage}
        & 
        \begin{minipage}[c][5em][c]{5.2cm}
        \centering
        Equation~(\ref{eq:pla_full})
        \end{minipage}
        & 
        \begin{minipage}[c][5em][c]{5.2cm}
        \centering
        Equation~(\ref{eq:cyl_full})
        \end{minipage}
        & 
        \begin{minipage}[c][5em][c]{5.2cm}
        \centering
        Equation~(\ref{eq:sph_full})
        \end{minipage}
        \\
        \hline 
        \begin{minipage}[c][5em][c]{1.5cm}
        $\tau_{\rm s} = 0$\\ $x_{\rm i} = 0$\\$v_{\rm E}/b \neq 0$\\Fit 
        \end{minipage}
        &
        \begin{minipage}[c][5em][c]{5.2cm}
        \centering
        Equation~(\ref{eq:ana_central_ext_pla}) and \\
        Table~\ref{tab:fitfunc_geo} column Slab
        \end{minipage}
        &
        \begin{minipage}[c][5em][c]{5.2cm}
        \centering
        Equation~(\ref{eq:ana_central_ext}) and \\
        Table~\ref{tab:fitfunc_geo} column Cylindrical
        \end{minipage}
        &
        \begin{minipage}[c][5em][c]{5.2cm}
        \centering
        Equation~(\ref{eq:ana_central_ext_sph}) and \\
        Table~\ref{tab:fitfunc_geo} column Spherical
        \end{minipage}
        \\
        \hline
        \begin{minipage}[c][5em][c]{1.5cm}
        $\tau_{\rm s} \neq 0$\\ $x_{\rm i} \neq 0$\\$v_{\rm E}/b \neq 0$\\Series 
        \end{minipage}
        &
        \begin{minipage}[c][5em][c]{5.2cm}
        \centering
        Equation~(\ref{eq:sesv_full_sla})
        \end{minipage}
        &
        \begin{minipage}[c][5em][c]{5.2cm}
        \centering
        Equation~(\ref{eq:sesv_full})
        \end{minipage}
        &
        \begin{minipage}[c][5em][c]{5.2cm}
        \centering
        Equation~(\ref{eq:sesv_full_sph})
        \end{minipage}
        \\
		\hline\hline
        \begin{minipage}[c][5em][c]{1.5cm}
        Recoil effect
        \end{minipage}
        &
        \multicolumn{3}{c}{
         $J_{\rm r}=D J \exp(-\alpha x/x_{\rm T})$,
         with $\alpha=0.78$ and $D=\int_{-\infty}^{+\infty} {\rm d}x J \left/ \int_{-\infty}^{+\infty} {\rm d}x J \exp(-\alpha x/x_{\rm T})\right.$
         } 
        \\
        &
        \multicolumn{3}{c}{
         $J_{\rm r}$ -- spectrum with recoil, $J$ -- spectrum without recoil 
         }
        \\
        \hline
    \end{tabular}
    \raggedright
    {\it Note.} The dimension of the mean intensity $J$ is different among different geometries, which is caused by the normalisation choice of emissivity $j_\nu$ as discussed in Section~\ref{subsec:ses_cyl}. 
    All analytical formulae are obtained under the assumption of large optical depth. 
    The fitting formulae for $v_{\rm E}/b \neq 0$ cases have a limited applicable range, spanning $\log(v_{\rm E}/b)=(-1, 2)$ and $\log (a_{\rm v}\tau_0) = (2.2, 4.2)$ (i.e. $\log \tau_0=(4, 6)$ at $T=10\, \rm K$), set by the range of model parameters whose spectra we choose to fit. 
    The series solutions for $v_{\rm E}/b \neq 0$ cases have a limited applicable range as well, $|v_{\rm E}/b| \ll (a_{\rm v}\tau_0)^{2/3}/12$, set by the assumption when deriving $\Lya$ RT equations under constant velocity gradients \citep{Nebrin_2025}. 
    In the analytical function of the first row and column Cylindrical, the coefficients in front of the exponential functions are obtained by setting $\tau_{\rm s}=0$ and $x_{\rm i}=0$ in equation~(\ref{eq:cyl_full}). 
\end{table*}

For convenient usage, we summarize the analytical formulae of $\Lya$ spectra studied in this paper in Table~\ref{tab:summary_formulae}. 
We will also provide the Python3 package for evaluating these analytical formulae on \url{https://github.com/PengfeiLiAstro/LyaRTAnalytical.git}. 
%%%%%%%%%%%%%%%%%%%%%%%%%%%%%%%%%%%%%%%%%%%%%%%%%%

% Don't change these lines
\bsp	% typesetting comment
\label{lastpage}
\end{document}